\documentclass[%
 reprint,
 amsmath,amssymb,
 aps,
]{revtex4-2}

\usepackage{braket}
\usepackage{graphicx}
\usepackage{dcolumn}
\usepackage{bm,upgreek}

\renewcommand\bra[1]{{\langle{#1}|}}
\makeatletter
\renewcommand\ket[1]{%
  \@ifnextchar\bra{\k@t{#1}\!}{\k@t{#1}}%
}
\newcommand\k@t[1]{{|{#1}\rangle}}
\makeatother

\begin{document}

\preprint{APS/123-QED}

\title{Quantum limits to resolution and discrimination of spontaneous emission lifetimes}

\author{Cheyenne S. Mitchell}
\author{Mikael P. Backlund}%
 \email{mikaelb@illinois.edu}
\affiliation{%
 Department of Chemistry, University of Illinois at Urbana-Champaign, Urbana, IL 61801
}%
\affiliation{%
 Illinois Quantum Information Science and Technology Center (IQUIST), University of Illinois at Urbana-Champaign, Urbana, IL 61801
}%
\date{\today}

\begin{abstract}
In this work we investigate the quantum information theoretical limits to several tasks related to lifetime estimation and discrimination of a two-level spontaneous optical emitter. We focus in particular on the model problem of resolving two mutually incoherent exponential decays with highly overlapping temporal probability profiles. Mirroring recent work on quantum-inspired super-resolution of point emitters, we find that direct lifetime measurement suffers from an analogue of ``Rayleigh’s Curse’’ when the time constants of the two decay channels approach one another. We propose alternative measurement schemes that circumvent this limit, and also demonstrate superiority to direct measurement for a related binary hypothesis test. Our findings add to a growing list of examples in which a quantum analysis uncovers significant information gains for certain tasks in opto-molecular metrology that do not rely on multiphoton interference, but evidently do benefit from a more thorough exploitation of the coherence properties of single photons. 
\end{abstract}

\maketitle


\section{\label{sec:level1}Introduction\protect}
\subsection{\label{sec:level2}Background}

Measurement of the spontaneous emission lifetime of one or more single-photon emitters is a technique of central importance across the chemical and physical sciences. In the context of single-molecule (SM) microscopy, fluorescence lifetime is routinely used as a sensitive indicator of molecular-scale changes in the probe's environment \cite{moerner2003methods,lakowicz2013principles}. The degree of freedom that determines the time delay between excitation pulse and subsequent photodetection is thus a conduit of chemical and physical information. Moreover, since SM emitters typically exhibit a finite budget of photocycles before irreversibly photobleaching, and since nanoscale environmental fluctuations can occur on timescales comparable to the lifetime, this resource can be an especially precious one. A careful analysis is thus warranted as to how one might carry out the tasks of lifetime estimation and discrimination as efficiently as possible with respect to photon budget, and certain metrics from classical information theory are now routinely invoked in such SM analyses \cite{chao2016fisher}. 

\begin{figure}
    \centering
    \includegraphics{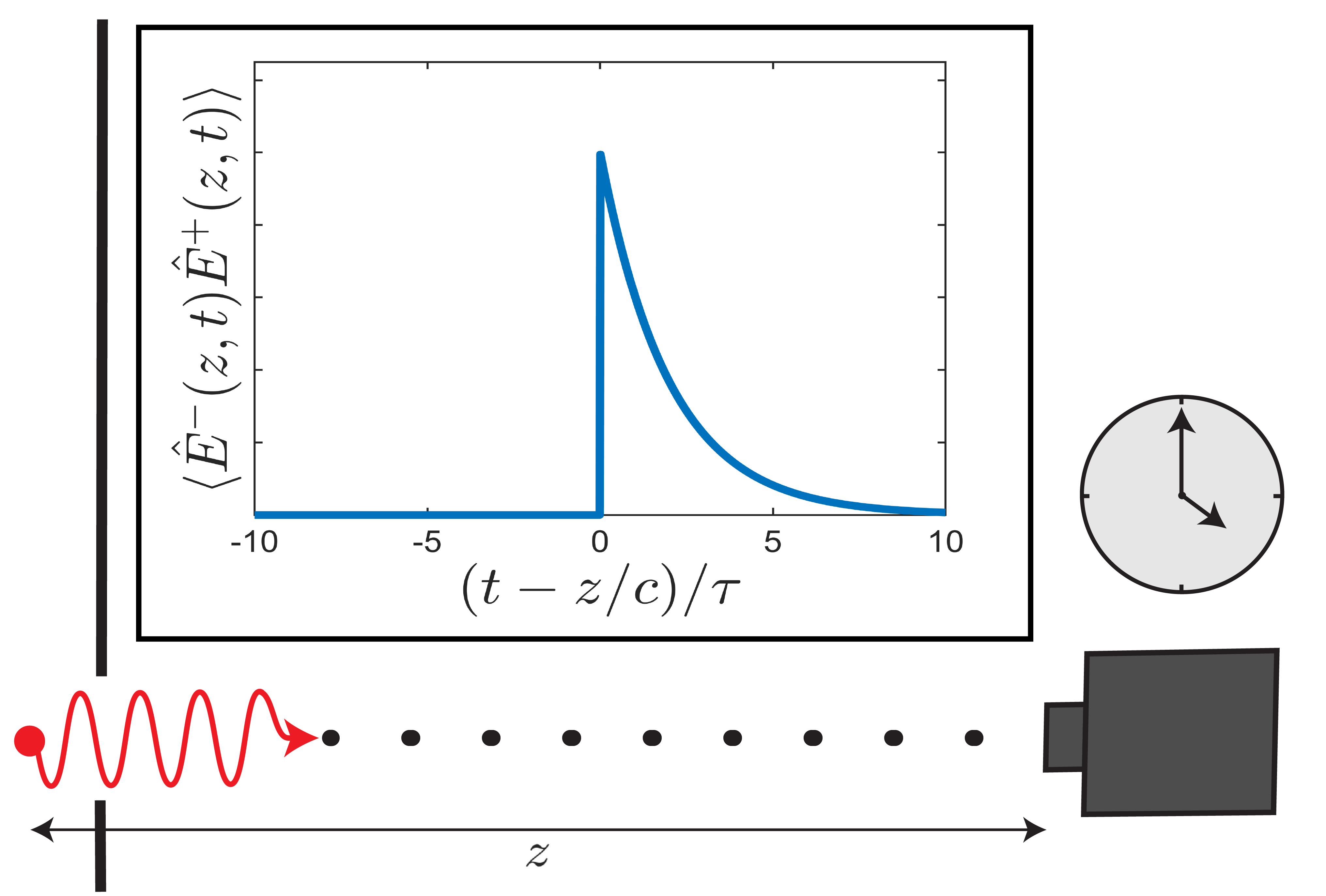}
    \caption{Direct measurement of fluorescence lifetime. A two-level emitter initially prepared in its excited state with the electromagnetic field in vacuum will eventually decay, leaving the field in a one-photon state. Assuming the resulting photon occupies modes within some accepted range, a detector placed some distance $z$ from the emitter has a probability of recording the photon at time $t$ proportional to the correlation function that labels the vertical axis of the inset \cite{scully_zubairy_1997}. In the long-time limit the probability profile is a one-sided exponential with a characteristic lifetime $\tau$. Direct measurement of lifetime is carried out by recording the time delay between excitation pulse and photodetection event, then repeating for many photons. The resulting histogram of arrival times can be fit to recover an estimate of $\tau$.}
    \label{fig:overview}
\end{figure}

In classical treatments of fluorescence lifetime measurement one typically models the photon arrival time as an exponential random variable with probability density:
\begin{equation} \label{eq_classical_lifetime_pdf}
    f(t;\tau) = \frac{1}{\tau} H(t) e^{-t/\tau}
\end{equation}
where $\tau$ is the fluorescence lifetime parameter and $H(t)$ is the Heaviside step function defined by $H(t)=0$ for $t<0$ and $H(t)=1$ for $t\geq0$. Note that the time origin is arbitrarily set to account for the timing of the preceding excitation pulse as well as the finite distance between emitter and detector. Multi-exponential decay can be modelled classically via a statistical mixture of probability densities of the form in Eq. (\ref{eq_classical_lifetime_pdf}) with different lifetimes $\{\tau_0, \tau_1, ...\}$. This parameterization allows one to compute the classical Fisher information (CFI) associated with estimation of $\{\tau_0, \tau_1,...\}$ (as well as the relative amplitudes of the mixture), as presented, e.g., in Ref. \cite{Bouchet:19}. The inverse of the CFI gives the classical Cram\'{e}r-Rao bound (CCRB), which sets the minimum variance of any unbiased estimator of these parameters. Thus the CFI and CCRB are useful metrics for assessing the efficiency of lifetime estimators given use of the Time-Correlated Single-Photon Counting (TCSPC) measurement apparatus depicted in Fig. \ref{fig:overview}, which we henceforth refer to as the ``direct measurement'' scheme. In practice estimator choice is only one optimization concern, and a more general optimization over all possible measurement schemes (i.e. other than direct measurement) provides more fundamental information bounds. Such a generalization of the CFI and CCRB are given by the quantum Fisher information (QFI) and quantum Cram\'{e}r-Rao bound (QCRB), which enumerate the information contained within the quantum state of the system (in this case the quantum state of the emitted photon) about the estimanda of interest, irrespective of the choice of Positive Operator-Valued Measure (POVM) \cite{helstrom1969quantum}. In this work we will present and analyze the QFI, QCRB, and related quantum information metrics for some illustrative, challenging tasks related to lifetime measurement of spontaneous emission.

In particular, the majority of our discussion will be devoted to the representative task of resolving a bi-exponential decay process via estimation of the ``separation'' (in a generalized sense, \textit{vide infra}) of the lifetime parameters. This effort is inspired in part by recent theoretical and experimental work that demonstrates the power in applying quantum information metrics to the analogous problem of \textit{spatial} resolution between mutually incoherent point sources \cite{doi:10.1080/00107514.2020.1736375,tsang2016quantum,PhysRevLett.117.190801,Nair:16,Tsang_2017,PhysRevA.95.063847,PhysRevA.97.023830,Rehacek:17,Tang:16,Larson:19,Hassett:18,Zhou:19,PhysRevLett.118.070801,Paur:16,Paur:18,Paur:19,PhysRevA.104.022613}. Namely, the CCRB associated with estimation of the source separation as measured by direct imaging is known to diverge at small separations (deemed ``Rayleigh's Curse''), while the corresponding QCRB remains finite. The discrepancy can be bridged by implementing a more sophisticated measurement scheme, e.g. Spatial-Mode Demultiplexing (SPADE) \cite{tsang2016quantum,Rehacek:17}, in order to recover otherwise wasted information. The present study is also related to recent applications of SPADE-like measurement techniques to the temporal resolution of shaped ultrafast optical pulses \cite{PhysRevLett.121.090501,PhysRevResearch.3.033082,PRXQuantum.2.010301}. Our work is distinguished by the facts that we consider the distinct pulse shapes (exponential) and timings ($\sim$nanoseconds) relevant for spontaneous optical emission, and by the fact that our resolution of lifetimes refers to the difference in temporal line \textit{widths} rather than the offset between peaks of subsequent pulses.  
\subsection{\label{sec:level2}Model}
We model the emitter as a two level system located at the origin with energy splitting $\hbar \omega_0$ and consider the evolution of the coupled emitter-electromagnetic field system after initially preparing the emitter in the excited state with the quantized electromagnetic field in the vacuum state. We invoke the Weisskopf-Wigner approximation and consider the long-time limit such that at some time $t$ later the emitter has returned to the ground state and the electromagnetic field is left in the one-photon state \cite{scully_zubairy_1997,novotny2012principles}:
\begin{equation} \label{eq_WW_photonstate}
    \ket{\psi_\tau} = \sum_{\lambda}\int \mathrm{d}^3\mathbf{k} \frac{g(\mathbf{k},\lambda)}{(\omega_\mathbf{k}-\omega_0) + i/(2\tau)} a^\dagger_{\mathbf{k},\lambda}\ket{0},
\end{equation}
where the index $\lambda$ denotes the polarization of the field mode, $\mathbf{k}$ is the linear momentum, $g(\mathbf{k},\lambda)$ subsumes relevant constants as well as the orientational factor associated with electric dipole emission, and $a^\dagger_{\mathbf{k},\lambda}$ is the canonical bosonic creation operator for the corresponding mode. For our purposes we ignore the Lamb shift to the photon's mean frequency. We can simplify the relevant one-photon state by imagining placing a detector with finite aperture some distance away such that only modes within some small solid angle avail themselves to photodetection. Let's also consider just the one linear polarization mode that dominates in the direction of the detector given the dipolar Green's tensor \cite{novotny2012principles}. Tracing out the unimportant modes (and ignoring the information-less vacuum contribution to the state of the field) leaves the normalized one-photon state:
\begin{equation} \label{eq_1photfreq}
    \ket{\psi_\tau} = \int_{-\infty}^\infty \mathrm{d}\omega \frac{1/\sqrt{2 \pi \tau}}{ (\omega-\omega_0) + i/(2\tau)} a^\dagger(\omega) \ket{0}
\end{equation}
where we have made use of the fact that $\omega_0 \tau \gg 0$ to extend the lower limit of the integral to $-\infty$. We can alternatively express $\ket{\psi_{\tau}}$ in terms of the Fourier-transformed creation operator \cite{loudon2000quantum} (up to a global phase):
\begin{equation} \label{eq_1phottime}
    \ket{\psi_\tau} = \int_{-\infty}^\infty \mathrm{d}t \frac{H(t)}{\sqrt{\tau}} e^{-i \omega_0 t} e^{-\frac{t}{2\tau}} a^\dagger(t) \ket{0},
\end{equation}
wherein $[a(t),a^\dagger(t')]=\delta(t-t')$. The Fourier integral in going from Eq. (\ref{eq_1photfreq}) to Eq. (\ref{eq_1phottime}) can be computed via complex contour integration. Define the function:
\begin{equation} \label{eq_psi_defn}
    \psi(t;\tau) \equiv \frac{H(t)}{\sqrt{\tau}} e^{-i \omega_0 t} e^{-\frac{t}{2\tau}}
\end{equation}
such that
\begin{equation} \label{eq_1phot_psi}
    \ket{\psi_\tau} = \int_{-\infty}^\infty \mathrm{d}t \, \psi(t;\tau) a^\dagger(t) \ket{0}.
\end{equation}
We will use Eq. (\ref{eq_1phot_psi}) as a starting point for the computation of quantum information metrics described in the remainder of the manuscript. Note that $\langle a^\dagger(t)a(t) \rangle = |\psi(t;\tau)|^2$, which coincides with the classical probability density function $f(t;\tau)$ as defined in Eq. (\ref{eq_classical_lifetime_pdf}).

If the two-level system is embedded in free space the relation between lifetime $\tau$ and center frequency $\omega_0$ is fixed via:
\begin{equation}
    \tau_{\text{free}} = \frac{3 \pi \epsilon_0 \hbar c^3}{\omega_0^3 \mu^2}
\end{equation}
where $\epsilon_0$ is the permittivity of free space and $\mu$ is the magnitude of the emitter's transition dipole moment \cite{scully_zubairy_1997,novotny2012principles}. For an emitter embedded in a dielectric medium, however, the relationship is altered in a way that depends explicitly on the (possibly inhomogeneous) dielectric constant \cite{novotny2012principles}. We focus our discussion by considering the case in which $\omega_0$ is fixed (and known) but $\tau$ is variable, e.g. in response to a changing local environment in the vicinity of the emitter.

\section{\label{sec:level1}Results and Discussion\protect}
\subsection{\label{sec:level2}Classical and quantum bounds of single-exponential lifetime estimation}
The QFI matrix $\mathcal{K} \in \mathbb{R}^{k \times k}$, the CFI matrix $\mathcal{J} \in \mathbb{R}^{k \times k}$, and the covariance matrix $\Sigma \in \mathbb{R}^{k \times k}$ of an arbitrary unbiased estimator for a given set of $k$ estimanda are related via:
\begin{equation}
    \Sigma \geq \mathcal{J}^{-1} \geq \mathcal{K}^{-1}
\end{equation}
where the matrix inequalities are meant to indicate that $(\mathcal{J}^{-1}-\mathcal{K}^{-1})$ and $(\Sigma-\mathcal{J}^{-1})$ are positive semidefinite. We begin by considering an emitter decaying with a single lifetime $\tau$ to be estimated. In this case the CFI and QFI are both scalars. For calculation of the CFI we specify a direct measurement in which an excitation pulse that is much shorter than the lifetime illuminates the system, then a subsequent photodetection event is tagged by its arrival time. In practice the time-tagging must be binned, but we consider the limiting case of infinitesimally short time bins for the calculation of $\mathcal{J}$. If no photodetection event occurs then the datum is simply thrown out. The cycle may be repeated until $N_\text{photons}$ photodetections have been recorded. The per-photon CFI associated with estimation of $\tau$ from such a measurement can be computed:
\begin{equation}
    \mathcal{J}_\tau^\text{(direct)} = \int_{0}^\infty \mathrm{d}t \frac{\left[ \frac{\partial f(t;\tau)}{\partial \tau} \right]^2}{f(t;\tau)} = \frac{1}{\tau^2}.
\end{equation}
The CFI associated with $N_\text{photons}$ such photodetections is then simply $N_\text{photons} \times \mathcal{J}_\tau^\text{(direct)}$.

Next we compute the QFI associated with estimating $\tau$ given the pure state $\ket{\psi_\tau}$, irrespective of the measurement scheme. Since the state is pure, we can compute the per-photon QFI directly via the Fubini-Study metric \cite{liu2019quantum}:
\begin{equation}
    \mathcal{K}_\tau = 4 \text{Re} \left( \braket{\partial_\tau \psi_\tau|\partial_\tau \psi_\tau} - \left| \braket{\partial_\tau \psi_\tau|\psi_\tau} \right|^2\right)
\end{equation}
with:
\begin{equation}
    \ket{\partial_\tau \psi_\tau} = \int_{-\infty}^\infty \mathrm{d}t \left[\frac{\partial \psi(t;\tau)}{\partial \tau}\right] a^\dagger(t) \ket{0}.
\end{equation}
The calculation is straightforward and yields:
\begin{equation}
    \mathcal{K}_\tau = \frac{1}{\tau^2}
\end{equation}
and so we conclude that in this case the CFI associated with direct measurement saturates the QFI. Direct measurement is optimal for this particular estimation task. More interesting cases are encountered when we consider multi-exponential decays due to incoherent mixtures of single-exponential processes.
\subsection{\label{sec:level2}Classical and quantum bounds of symmetric bi-exponential lifetime resolution}
\begin{figure}
    \centering
    \includegraphics{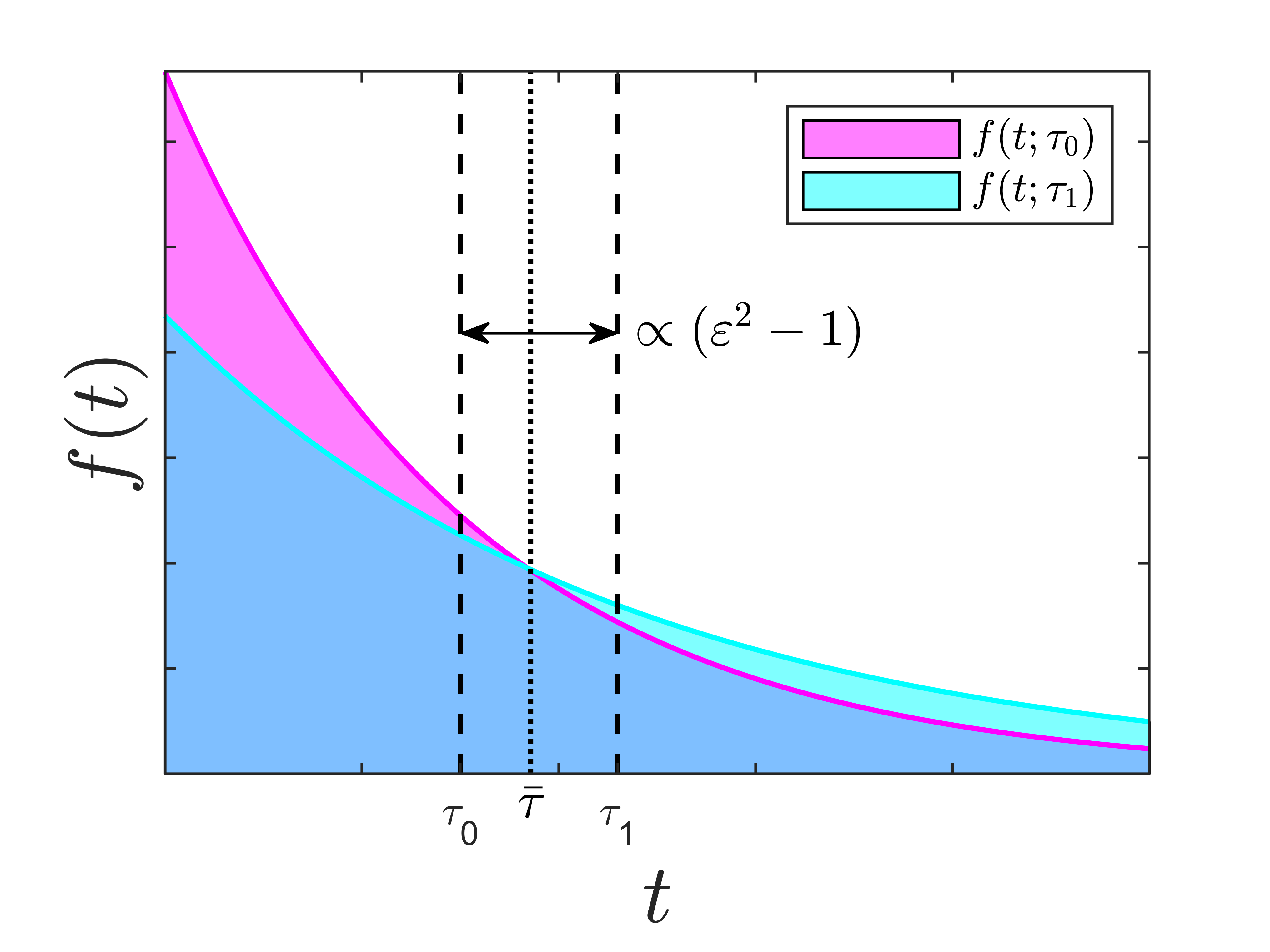}
    \caption{Illustration of basic lifetime resolution task. The classical probability distribution resulting from decay with constants $\tau_0$ and $\tau_1$ (magenta and cyan, respectively). Blue region marks overlap between the distributions. Dashed vertical lines mark the lifetimes along the time axis, while the dotted line marks their geometric mean $\bar{\tau}$. The difference between the lifetimes is proportional to $\varepsilon^2-1$.}
    \label{fig:resolution_scheme}
\end{figure}
\begin{figure}
    \centering
    \includegraphics{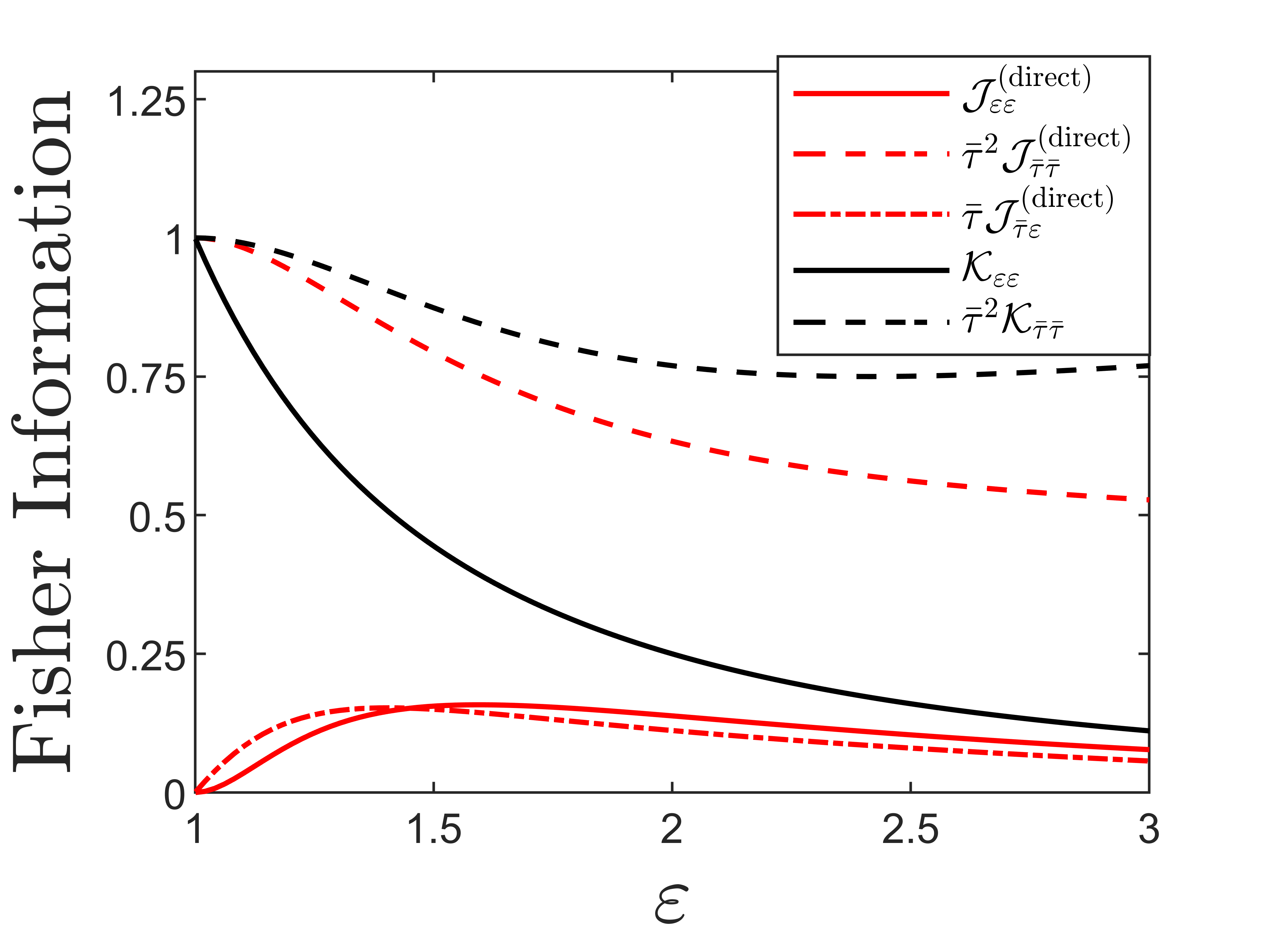}
    \caption{Per-photon Fisher information associated with estimation of $\bar{\tau}$ and $\varepsilon$. Red lines correspond to the elements of the direct measurement CFI matrix, while the black lines indicate the nonzero elements of the QFI matrix. Notably, $\mathcal{J}^\text{(direct)}_{\varepsilon \varepsilon}$ vanishes as $\varepsilon \to 1$, while $\mathcal{K}_{\varepsilon \varepsilon}$ takes its maximum value in the same limit.}
    \label{fig:direct_vs_quantum_FI}
\end{figure}

\begin{figure}
    \centering
    \includegraphics{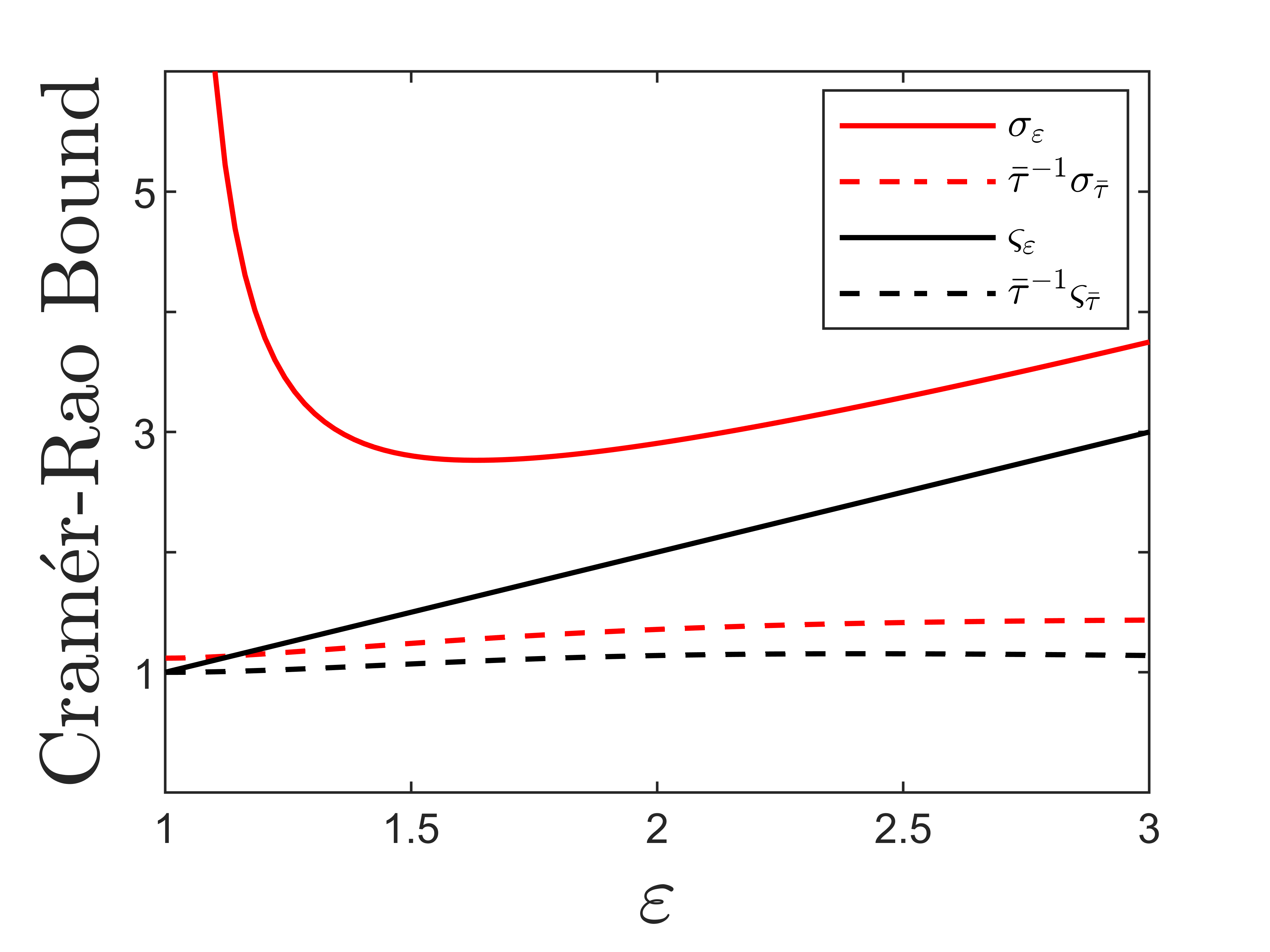}
    \caption{Scaled (square-root) Cram\'{e}r-Rao bounds associated with estimation of $\bar{\tau}$ and $\varepsilon$. Red lines give the classical bounds for a direct measurement, while black lines indicate the quantum bounds. Notably, $\sigma_\varepsilon$ diverges as $\varepsilon \to 1$, while $\varsigma_\varepsilon$ assumes its minimum value in the same limit. }
    \label{fig:direct_vs_quantum_CRB}
\end{figure}

The most basic such case is that in which the relevant single-photon state is an equal-probability mixture of two single exponential decays with lifetimes $\tau_0$ and $\tau_1$ (Fig. \ref{fig:resolution_scheme}) such that:
\begin{equation} \label{eq_rho_resolution}
    \rho = \frac{1}{2}\ket{\psi_0}\bra{\psi_0} + \frac{1}{2}\ket{\psi_1}\bra{\psi_1}
\end{equation}
with:
\begin{subequations}
\begin{equation}
    \ket{\psi_0} = \int_{-\infty}^\infty \mathrm{d}t \, \psi(t;\tau_0) a^\dagger(t) \ket{0}
\end{equation}
\begin{equation}
    \ket{\psi_1} = \int_{-\infty}^\infty \mathrm{d}t \, \psi(t;\tau_1) a^\dagger(t) \ket{0}.
\end{equation}
\end{subequations}
Without loss of generality we suppose $\tau_1 \geq \tau_0$. The FI matrices associated with estimating $\tau_0$ and $\tau_1$ are of dimensions $2\times2$. Rather than considering estimation of $\tau_0$ and $\tau_1$ directly, we re-parameterize the task into the estimation of the geometric mean lifetime, $\bar{\tau}$, and the square-root ratio, $\varepsilon$, in order to highlight the fact that it's really the latter that poses a fundamental challenge for direct measurement. These coordinate-transformed estimanda are defined by:
\begin{subequations}
\begin{equation}
    \bar{\tau} = \sqrt{\tau_0 \tau_1}
\end{equation}
\begin{equation}
    \varepsilon = \sqrt{\frac{\tau_1}{\tau_0}}
\end{equation}
\end{subequations}
such that $\tau_0 = \bar{\tau}/\varepsilon$ and $\tau_1 = \varepsilon \bar{\tau}$. Our choice of $\tau_1 \geq \tau_0$ guarantees $\varepsilon \geq 1$. Let the vector of parameters be defined $\bm{\uptheta}=(\bar{\tau},\varepsilon)^\text{T}$. We first consider the CFI with respect to direct measurement. The relevant classical probability density function is:
\begin{equation}
    f_{\rho}(t;\bm{\uptheta}) = \frac{1}{2}f(t;\bar{\tau}/\varepsilon) + \frac{1}{2}f(t;\varepsilon \bar{\tau})
\end{equation}
with $f(t;\tau)$ defined as in Eq. (\ref{eq_classical_lifetime_pdf}). The elements of $\mathcal{J}^\text{(direct)}$ can be computed from:
\begin{equation}
        \mathcal{J}_{ij}^\text{(direct)} = \int_{0}^\infty \mathrm{d}t \frac{\left[ \frac{\partial f_\rho(t;\bm{\uptheta})}{\partial \theta_i} \right] \left[ \frac{\partial f_\rho(t;\bm{\uptheta})}{\partial \theta_j} \right]}{f_\rho(t;\bm{\uptheta})}.
\end{equation}
The resulting matrix is:
\begin{equation}
    \mathcal{J}^\text{(direct)} = 
    \begin{pmatrix} 
    \mathcal{J}_{\bar{\tau}\bar{\tau}}^\text{(direct)} & \mathcal{J}_{\bar{\tau}\varepsilon}^\text{(direct)} \\
    \mathcal{J}_{\bar{\tau}\varepsilon}^\text{(direct)} & \mathcal{J}_{\varepsilon\varepsilon}^\text{(direct)}
    \end{pmatrix}.
\end{equation} 
\vspace{1mm}
\noindent We compute $\mathcal{J}_{\bar{\tau}\bar{\tau}}^\text{(direct)}$, $\mathcal{J}_{\bar{\tau}\varepsilon}^\text{(direct)}$, and $\mathcal{J}_{\varepsilon\varepsilon}^\text{(direct)}$ numerically and plot their scaled ($\bar{\tau}$-independent) values as functions of $\varepsilon$ in Fig. \ref{fig:direct_vs_quantum_FI}. Notably, while $\mathcal{J}_{\bar{\tau}\bar{\tau}}^\text{(direct)}$ peaks as $\varepsilon \rightarrow 1$, $\mathcal{J}_{\varepsilon\varepsilon}^\text{(direct)}$ vanishes in the same limit. We plot appropriately scaled square-root CCRB values in Fig. \ref{fig:direct_vs_quantum_CRB} defined by:
\begin{subequations}
\begin{equation}
    \sigma_{\bar{\tau}} = \sqrt{[(\mathcal{J}^\text{(direct)})^{-1}]_{11}}
\end{equation}
\begin{equation}
    \sigma_{\varepsilon} = \sqrt{[(\mathcal{J}^\text{(direct)})^{-1}]_{22}}.
\end{equation}
\end{subequations}
It's evident that $\sigma_{\varepsilon}$ diverges as $\varepsilon \rightarrow 1$. These observations are analogous to the onset of ``Rayleigh's Curse'' as described in Ref. \cite{doi:10.1080/00107514.2020.1736375}, wherein it becomes increasingly difficult to produce a precise, unbiased estimate of the spatial separation between two point sources as they are brought closer together.
Next we turn our attention to computing the QFI with respect to $\bar{\tau}$ and $\varepsilon$. Since $\rho$ represents a mixed state we must compute the elements of $\mathcal{K}$ via the more general \cite{PhysRevLett.72.3439}:
\begin{equation} \label{eq_QFI_defn}
    \mathcal{K}_{ij} = \text{Re Tr } \left(\mathcal{L}_i\mathcal{L}_j\rho\right)
\end{equation}
where $\mathcal{L}_i$ is the symmetric logarithmic derivative (SLD) operator associated with the parameter $\theta_i \in \{\bar{\tau},\varepsilon \}$, defined implicitly via:
\begin{equation} \label{eq_SLD_implicit}
    \partial_{\theta_i}\rho = \frac{1}{2}\left( \rho \mathcal{L}_i + \mathcal{L}_i\rho \right).
\end{equation}
Details of the calculation of $\mathcal{K}$ can be found in Appendix \ref{appendix_A}. The result is:
\begin{equation} \label{eq_QFI_resolution}
    \mathcal{K} = \begin{pmatrix}
    \frac{1+14\varepsilon^4 + \varepsilon^8}{(1+\varepsilon^2)^4\bar{\tau}^2} && 0 \\ 0 && \frac{1}{\varepsilon^2}
    \end{pmatrix}.
\end{equation}
Note that $\mathcal{K}_{\bar{\tau}\bar{\tau}} \rightarrow 1/\bar{\tau}^2$ as $\varepsilon \rightarrow 1$. More importantly, we find that $\mathcal{K}_{\varepsilon\varepsilon}$ not only remains nonzero as $\varepsilon \rightarrow 1$, it actually takes on it's maximum value in this limit. We compare the direct measurement CFI with the measurement-agnostic QFI in Fig. \ref{fig:direct_vs_quantum_FI}. The square-root QCRBs are defined by:
\begin{subequations}
\begin{equation}
    \varsigma_{\bar{\tau}} = \sqrt{[\mathcal{K}^{-1}]_{11}}
\end{equation}
\begin{equation}
    \varsigma_{\varepsilon} = \sqrt{[\mathcal{K}^{-1}]_{22}}.
\end{equation}
\end{subequations}
and are plotted vs. the direct measurement square-root CCRBs in Fig. \ref{fig:direct_vs_quantum_CRB}. Evidently direct measurement leaves an increasing fraction of available information on lifetime separation on the table as the two time constants $\tau_0$ and $\tau_1$ approach one another. In the regime where direct measurement has the most difficulty discerning the decay constants, we next show that there exists some alternative measurement scheme that performs optimally.
\subsection{\label{sec:level2}Optimal and near-optimal measurement schemes for symmetric bi-exponential lifetime resolution}
\subsubsection{Weighted Laguerre mode projection}
\begin{figure}
    \centering
    \includegraphics{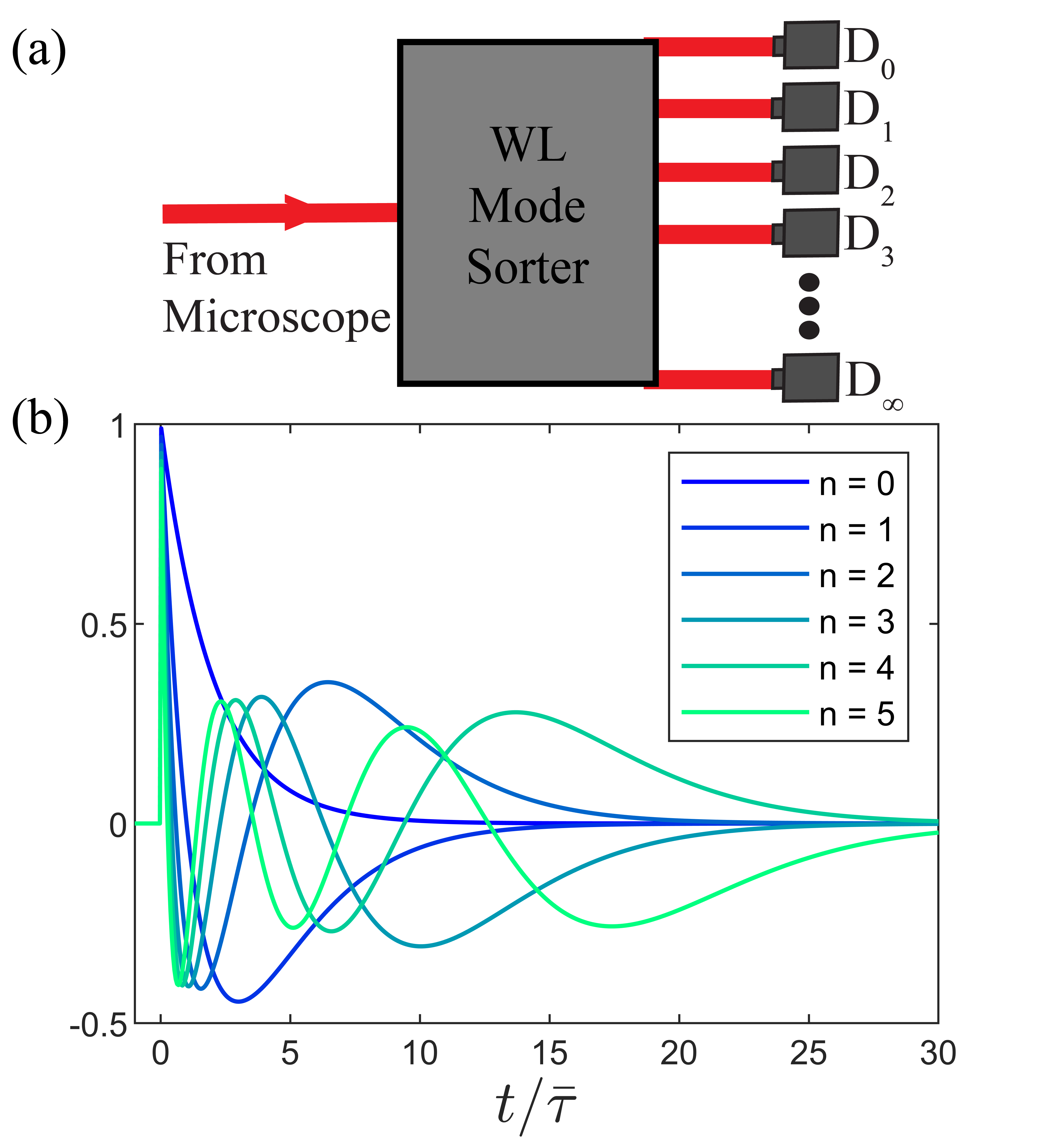}
    \caption{(a) Schematic of a mode sorting device that projects collected photons onto the basis of exponentially-weighted Laguerre polynomials. The detector D$_\infty$ records projection onto the complement of the space spanned by the finite number of modes recorded in the other channels. (b) Plots of the envelopes of the first few $\phi_n$ functions. The high frequency component $e^{-i\omega_0t}$ is suppressed in each for clarity.}
    \label{fig:WL_sorter}
\end{figure}

In Ref. \cite{tsang2016quantum} Tsang and coworkers showed that for the task of estimating the spatial separation between two incoherent point emitters subject to a Gaussian point spread function (PSF), direct imaging fails for small separation while a measurement that projects onto the Hermite-Gaussian spatial modes performs optimally. Reference \cite{Rehacek:17} generalizes this basic result by showing that projection onto a complete set of modes with definite parity is optimal for any symmetric PSF, and that projection onto an orthonormalized set of polynomials related to the derivatives of the amplitude PSF is efficient for the task. Here we report an analogous finding for the task of resolving the time constants of two incoherent and overlapping exponential decays. In particular, projection onto exponentially-weighted Laguerre polynomials achieves optimality (Fig. \ref{fig:WL_sorter}). The connection to the spatial problem is notable, as Hermite polynomials are orthogonal over the domain of integration $(-\infty,\infty)$ with respect to a Gaussian weighting function, while the Laguerre polynomials are orthogonal over $(0,\infty)$ with respect to an exponential weighting function \cite{cai2007weighted}.
We define the orthonormal weighted Laguerre (WL) functions via:
\begin{equation} \label{eq_phi_defn}
    \phi_n(t;\bar{\tau}) = \frac{H(t)}{\sqrt{\bar{\tau}}} e^{-i \omega_0 t} e^{-\frac{t}{2 \bar{\tau}}} L_n\left(\frac{t}{\bar{\tau}}\right)
\end{equation}
where $L_n(\cdot)$ denotes the $n^\text{th}$ Laguerre polynomial for nonnegative integer $n$. The first few such functions are plotted in Fig. \ref{fig:WL_sorter}(b). Note that $\phi_0(t;\bar{\tau}) = \psi(t;\bar{\tau})$ as defined in Eq. (\ref{eq_psi_defn}). We construct the single-photon states:
\begin{equation}
    \ket{\phi_n(\bar{\tau})} = \int_{-\infty}^\infty \mathrm{d}t \, \phi_n(t) a^\dagger(t) \ket{0}
\end{equation}
associated with the projection measurement $\{\ket{\phi_n(\bar{\tau})}\bra{\phi_n(\bar{\tau})}\}_n$. Implementation of this measurement on the state $\rho$ gives an outcome with probability mass function:
\begin{equation}
    P_n(\varepsilon) = \frac{1}{2}|\braket{\phi_n(\bar{\tau})|\psi_0}|^2 + \frac{1}{2}|\braket{\phi_n(\bar{\tau})|\psi_1}|^2.
\end{equation}
Closed-form expressions for $\braket{\phi_n(\bar{\tau})|\psi_0}$ and $\braket{\phi_n(\bar{\tau})|\psi_1}$ are given by:
\begin{subequations}
\begin{equation}
    \braket{\phi_n(\bar{\tau})|\psi_0} = \frac{2 \left( \bar{\tau} - \tau_0 \right)^n \sqrt{\bar{\tau} \tau_0} }{\left( \bar{\tau} + \tau_0 \right)^{n+1}}
\end{equation}
\begin{equation}
    \braket{\phi_n(\bar{\tau})|\psi_1} = \frac{2 \left( \bar{\tau} - \tau_1 \right)^n \sqrt{\bar{\tau} \tau_1} }{\left( \bar{\tau} + \tau_1 \right)^{n+1}},
\end{equation}
\end{subequations}
which yields:
\begin{equation}
    P_n(\varepsilon) = 4\varepsilon \frac{(\varepsilon^2 - 1)^{2n}}{(\varepsilon+1)^{4n+2}}
\end{equation}
The CFI associated with estimation of $\varepsilon$ can be computed via:
\begin{equation}
    \mathcal{J}^\text{(WL)}_{\varepsilon\varepsilon} = \sum_{n=0}^\infty \frac{\left[\frac{\partial P_n}{\partial \varepsilon}\right]^2 }{P_n}
\end{equation}
to give
\begin{equation}
    \mathcal{J}^\text{(WL)}_{\varepsilon\varepsilon} = \frac{1}{\varepsilon^2} = \mathcal{K}_{\varepsilon\varepsilon}.
\end{equation}
Thus projection onto the WL modes saturates the QFI associated with estimation of $\varepsilon$. Note that in this treatment we have implicitly assumed that $\bar{\tau}$ is known \textit{a priori}, akin to the spatial resolution case in which optimality of the Hermite-Gaussian projection requires prior knowledge of the centroid \cite{tsang2016quantum}. Following the approach described in Ref. \cite{tsang2016quantum}, if $\bar{\tau}$ is not known we propose an adaptive measurement scheme in which $\bar{\tau}$ is first determined via direct measurement, then $\varepsilon$ is subsequently estimated from WL projections. In practice some finite uncertainty will remain in the initial estimate of $\bar{\tau}$. We explore the effect of the temporal equivalent of ``centroid misalignment'' in Appendix \ref{appendix_C}. 
\begin{figure}
    \centering
    \includegraphics{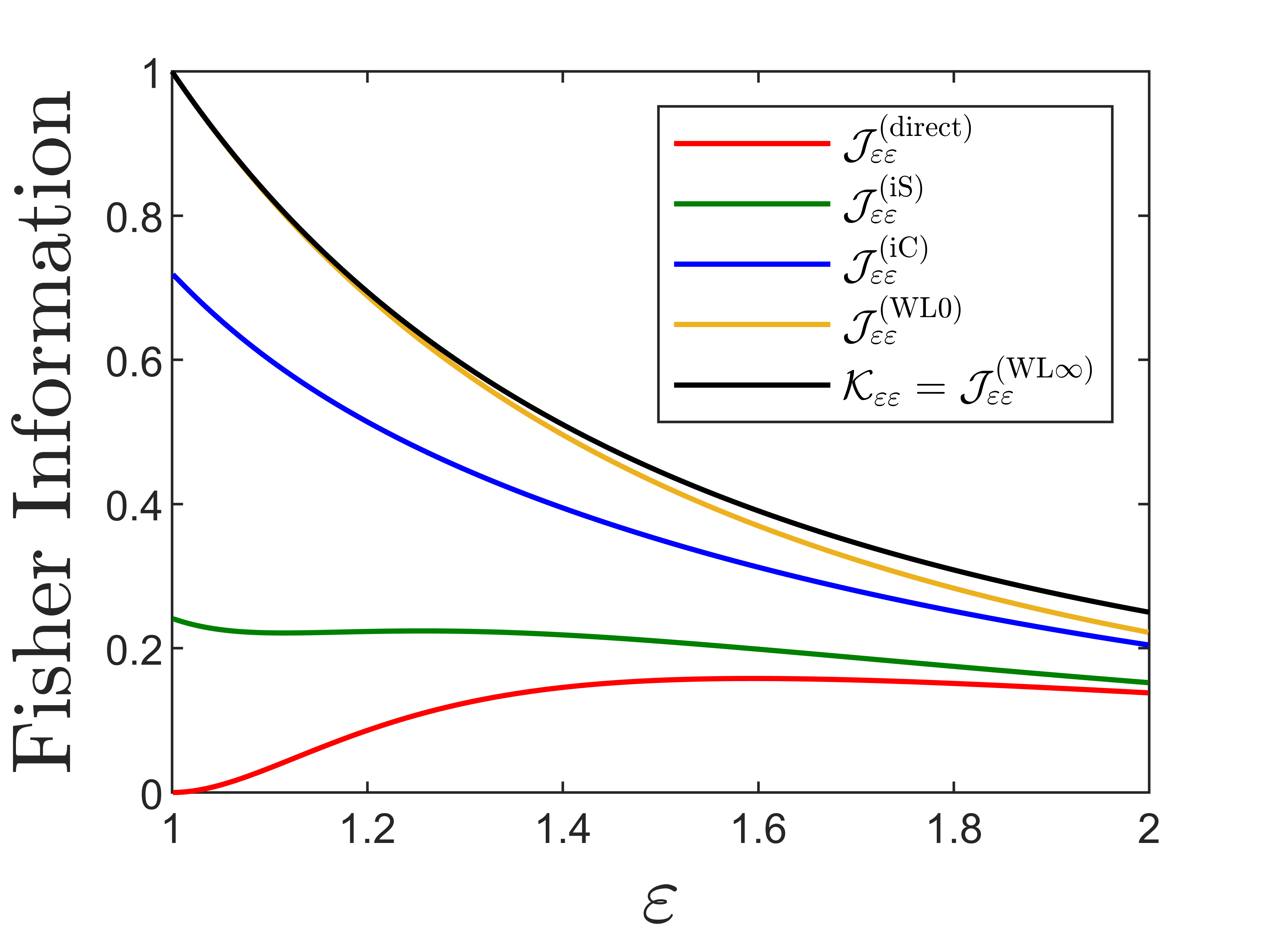}
    \caption{Comparison of the QFI associated with estimation of $\varepsilon$ to the CFI of various measurement schemes described in the text. The label ``WL0'' indicates the binary truncation of the weighted Laguerre mode sorter, while ``WL$\infty$'' indicates the limit in which infinitely many such modes are detected.}
    \label{fig:FI_comparison_various}
\end{figure}
We next speculate on how one might realize projection onto the WL modes with a physical measurement apparatus. Recent experimental studies have demonstrated quantum pulse gates as a means to sort mixtures of ultrafast optical pulses by their temporal mode composition \cite{PhysRevLett.121.090501,PhysRevResearch.3.033082,PRXQuantum.2.010301}. The method works by gating a target optical pulse/pulse train with a temporally shaped near-infrared control pulse. The target and control are passed together through a nonlinear medium (periodically poled lithium niobate) resulting in a frequency-shifted signal in proportion to their overlap. This effectively produces a projection measurement onto the mode of the control pulse \cite{Eckstein:11}. A similar approach may provide a route to experimental realization of the WL projections described above, though it should be noted that the $\sim$ns ``pulses'' produced by a spontaneous optical emitter are much longer than the ultrafast pulses treated in Refs. \cite{PhysRevLett.121.090501,PhysRevResearch.3.033082,PRXQuantum.2.010301}.

An interesting relation among the inverse Fourier transforms of the WL functions suggests another possible route to experimental realization. We take the following convention for the definition of the inverse Fourier transform of $\phi_n(t)$:
\begin{equation} \label{eq_define_phi_tilde}
    \tilde{\phi}_n(\omega) = \frac{1}{\sqrt{2\pi}} \int_{-\infty}^\infty \mathrm{d}t \, \phi_n(t) e^{i\omega t}.
\end{equation}
In Appendix \ref{appendix_B} we prove the relation:
\begin{equation} \label{eq_iFT_WL_relation}
    \tilde{\phi}_n(\omega) = \frac{i}{(\omega-\omega_0)\bar{\tau} + i/2}\sqrt{\frac{\bar{\tau}}{2\pi}} \exp\Big[{i n \Phi(\omega-\omega_0;\bar{\tau})}\Big],
\end{equation}
where $\Phi(\omega-\omega_0;\bar{\tau})$ is a real-valued function. In other words, each $\tilde{\phi}_n$ differs from the others in the sequence only in its phase, and:
\begin{equation}
    \tilde{\phi}_n(\omega) = \tilde{\phi}_{n-1}(\omega)\exp{\Big[i \Phi(\omega-\omega_0;\bar{\tau}) \Big]}
\end{equation}
for all $n>0$. If we can access the functions $\{\tilde{\phi}_n(\omega)\}_n$ then we can effect the transformation $\tilde{\phi}_n \to \tilde{\phi}_{n-1} \, \forall n>0$ by imparting a frequency-dependent phase delay of $-\Phi(\omega-\omega_0;\bar{\tau})$. This is suggestive of the technique of ultrafast pulse shaping, in which the temporal profile of short optical pulses is sculpted by passing through a 4f system consisting of first a dispersive element to map frequency into lateral position, followed by a spatial light modulator onto which the desired frequency-dependent phase modulation is encoded, and then a second dispersive element to recombine the spectral modes \cite{weiner2000femtosecond,weiner2011ultrafast}. In our case, however, we aim to reshape pulses of $\sim$ns temporal and $\sim$GHz frequency extent, a much different regime than the ultrafast, ultrabroadband pulses that are typically treated with this approach. Temporally reshaping such a narrow-band pulse would be exceedingly difficult without somehow initially compressing the pulse, or else taking a different route altogether \cite{rakher2011simultaneous,kielpinski2011quantum}. Nonetheless, the accompanying mathematics is intriguing enough to warrant a bit more speculation before moving on. 

The same phase modulation that transforms $\tilde{\phi}_n(\omega) \to \tilde{\phi}_{n-1}(\omega) \, \forall n>0$ also transforms $\tilde{\phi}_0(\omega) \to -\tilde{\phi}_0^*(\omega)$. Sandwiching this transformation between an inverse Fourier transform and Fourier transform implements the overall mapping $\phi_0(t) \to -\phi_0^*(-t)$, i.e. a reflection (up to a sign) of the temporal profile about the origin. In other words, the component of the signal in the zero mode would alone be advanced to early times, and so a simple time-gating measurement could pick off the projection onto this mode. One must recognize of course that causality should be maintained such that occupation of negative temporal modes is not indicative of superluminal propagation, but rather is an artifact of redefining $t=0$ to account for the propagation of the quiescent pulse through the apparatus \cite{weiner2011ultrafast}. If this time gating can be implemented with a temporal beam splitter then one can imagine implementing a sort of shift register that measures projection onto the $n^\text{th}$ WL mode by way of $n$ sequential phase modulations followed by time gating. We reserve a deeper investigation of this approach for future work.

One practical consideration of any experimental realization of WL mode projection is that the measurement must realistically be truncated at some finite $n$. Figure \ref{fig:FI_comparison_various} depicts the CFI associated with the binary measurement $\{\ket{\phi_0}\bra{\phi_0}, \mathbb{I}-\ket{\phi_0}\bra{\phi_0}\}$, indicating very nearly optimal performance for small $\varepsilon$. This is in analogy to the near-optimality of binary SPADE, as analyzed previously for the task of spatial resolution \cite{tsang2016quantum}.  

\subsubsection{Approximate interferometric approaches} \label{section_interfs}
\begin{figure}
    \centering
    \includegraphics{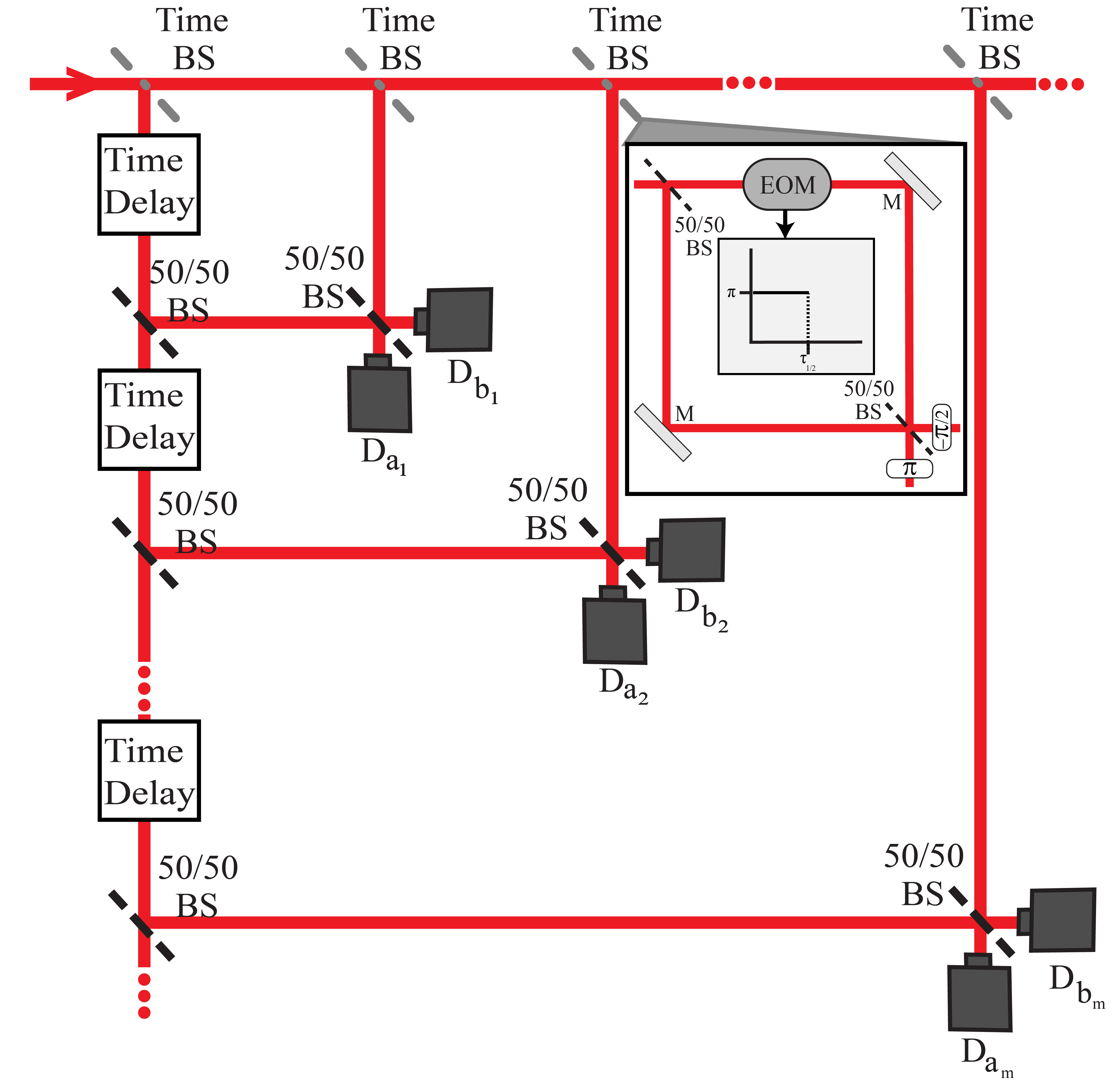}
    \caption{Schematic of proposed cascaded interferometer, ``iC'', with M: mirror, BS: beam splitter, D: detector. The time beam splitters along the top edge are implemented as described in the text and as indicated in the inset (EOM: electro-optic modulator). 50/50 beam splitters along the left edge are alternated with $\tau_{1/2}$-delay stages.}
    \label{fig:iC}
\end{figure}
We conclude this section by introducing and analyzing two interferometer arrangements that circumvent the lifetime analogue of ``Rayleigh's Curse''. First we consider the cascaded interferometer depicted in Fig. \ref{fig:iC} and heretofore referred to as the ``iC'' apparatus. This hypothetical scheme exploits the shift-and-scale symmetry of the exponential decay function given by:
\begin{equation}
    |\phi_0(t;\bar{\tau})|^2 = \frac{1}{\bar{\tau}} H(t) e^{-\frac{t}{\bar{\tau}}},
\end{equation}
which has half-life:
\begin{equation}
    \tau_{1/2} = \bar{\tau} \log{2}.
\end{equation}
We consider the transformation of the generic exponential amplitude function $\psi(t;\tau)$ [Eq. (\ref{eq_psi_defn})] after it enters the interferometer at the top left of Fig. \ref{fig:iC}. It first encounters a temporal beam splitter that acts to transmit $\psi(t)H(t-\tau_{1/2})$ and reflect $\psi(t)[H(t)-H(t-\tau_{1/2})]$. This could in theory be realized by a Mach-Zender interferometer (Fig. \ref{fig:iC} inset) in which a time-dependent phase flip is applied via a synchronized electro-optic modulator (EOM) placed in one arm. The transmitted portion then encounters a cascade of similar temporal beam splitters (top edge of Fig. \ref{fig:iC}), the $m^\text{th}$ of which reflects $\psi(t)\{H(t-m\tau_{1/2})-H(t-[m+1]\tau_{1/2})\}$ downward while transmitting $\psi(t)H(t-[m+1]\tau_{1/2})$ rightward to the subsequent stage. Meanwhile, the portion of the light sent downward along the leftmost edge of the interferometer by the first temporal beam splitter is subjected to a cascade of alternating delay stages (each set to delay by an additional $\tau_{1/2}$ relative to the top edge) and 50/50 beam splitters. The light reflected by the $m^\text{th}$ 50/50 beam splitter along the left edge has amplitude $i(\alpha/\sqrt{2})^m \psi(t)\{H(t-m\tau_{1/2})-H(t-[m+1]\tau_{1/2})\}$, where $\alpha = \exp{(\frac{\tau_{1/2}}{2 \tau})}$. The reflected outputs of the $m^\text{th}$ 50/50 beam splitter along the left edge and the $m^\text{th}$ temporal beam splitter along the top edge are recombined on another 50/50 beam splitter and the resulting interferogram is recorded at the $m^\text{th}$ detection stage. We numerically computed the CFI associated with $\varepsilon$ estimation using the iC scheme. In our calculation we effectively assume that we are free to fine-adjust phases equivalent to time delays on the order of $2\pi/\omega_0$ throughout the interferometer such that an overall input state of $\psi(t;\bar{\tau})$ would produce nulls in each of the ``a'' detection channels. The results are plotted in Fig. \ref{fig:FI_comparison_various} and indicate near-optimal performance that certainly overcomes the vanishing information associated with direct measurement. However, this interferometric cascade is admittedly complicated and would be extremely difficult to implement even for a few detection stages due to synchronization, alignment, and loss issues.
\begin{figure}
    \centering
    \includegraphics[width=4.3cm]{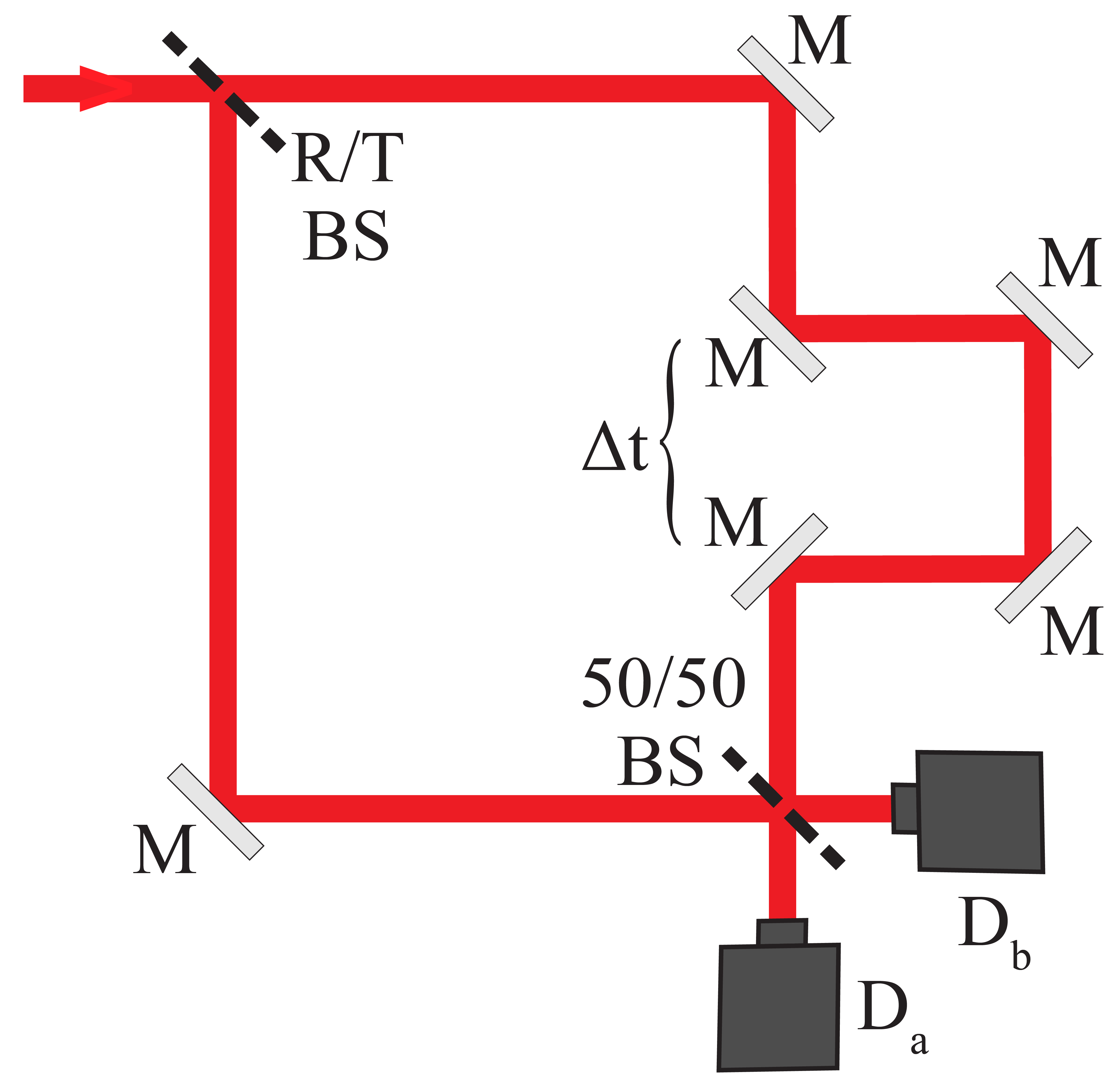}
    \caption{Schematic of proposed simplified interferometer, ``iS'', with $\Delta t \approx \bar{\tau} \log(R/T)$.}
    \label{fig:iS}
\end{figure}
It therefore behooves us to seek a simpler interferometric scheme that might not perform as efficiently in theory, but still successfully circumvents the lifetime analogue of ``Rayleigh's Curse''. 

Such a scheme is depicted in Fig. \ref{fig:iS}, which we refer to as ``iS''. Collected light enters iS through a beam splitter with reflectance $R = 0.9$ (complex amplitude $i\sqrt{0.9}$) and transmittance $T = 0.1$ (complex amplitude $\sqrt{0.1}$). The reflected portion is relayed to one input of a 50/50 beam splitter without further modulation. The initially transmitted portion is passed through a fixed delay stage producing a time lag of $\bar{\tau}\log(R/T)$ (plus some implied finely-tuned adjustment on the order of $2\pi/\omega_0$) , which is then sent to the other input port of the 50/50 beam splitter. Time-tagged photons are then counted in the two output channels. Numerical evaluation of the CFI associated with estimation of $\varepsilon$ using iS confirms finite information content as $\varepsilon \to 1$ (Fig. \ref{fig:FI_comparison_various}).

\subsection{\label{sec:level2}Weak decay path detection}
\begin{figure}
    \centering
    \includegraphics{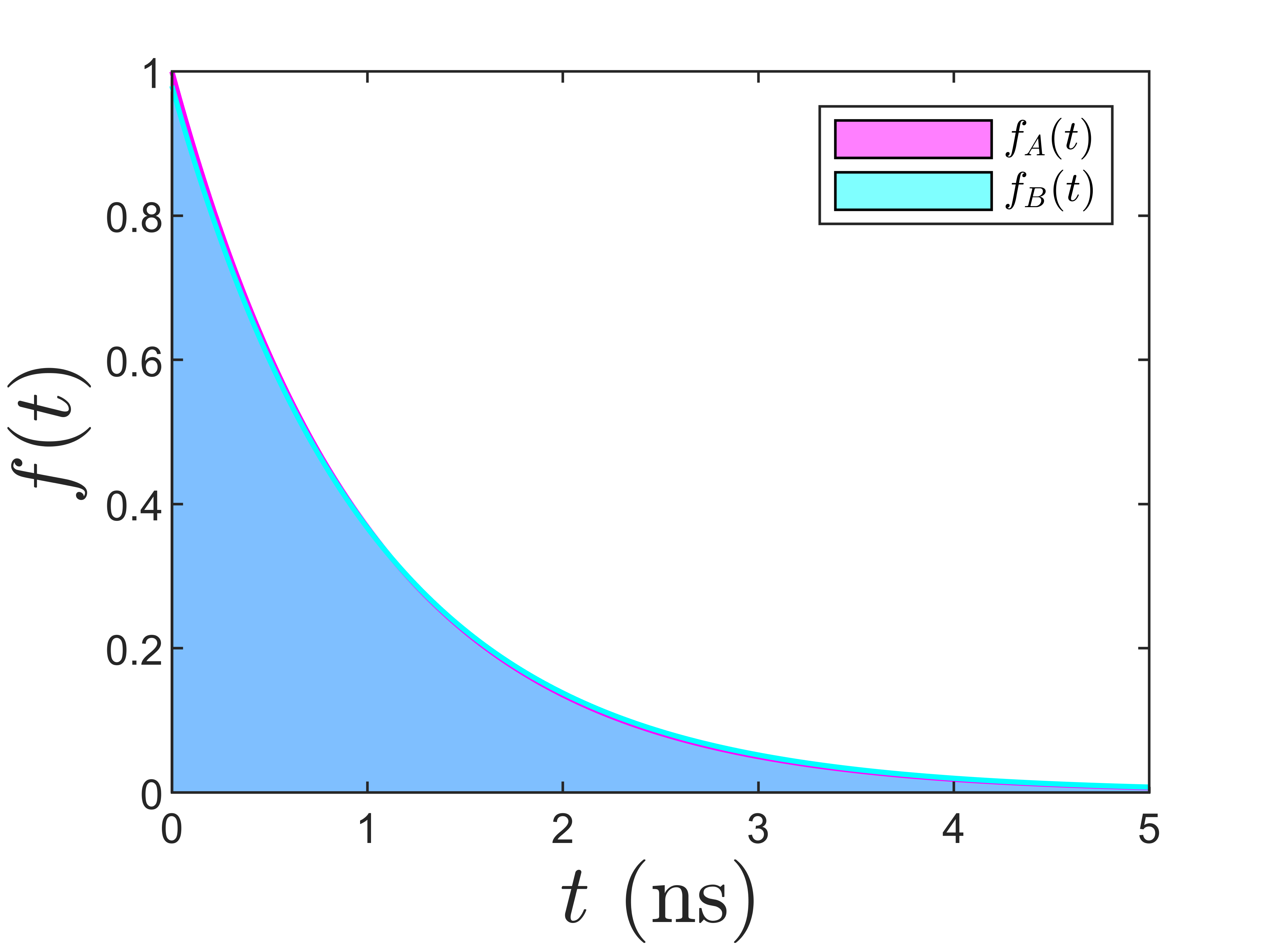}
    \caption{Illustration of weak decay path detection task. Classical probability distributions associated with States A (magenta) and B (cyan) to be discriminated. Blue marks the overlap of the distributions.}
    \label{fig:weakdecay_scheme}
\end{figure}
\begin{figure}
    \centering
    \includegraphics{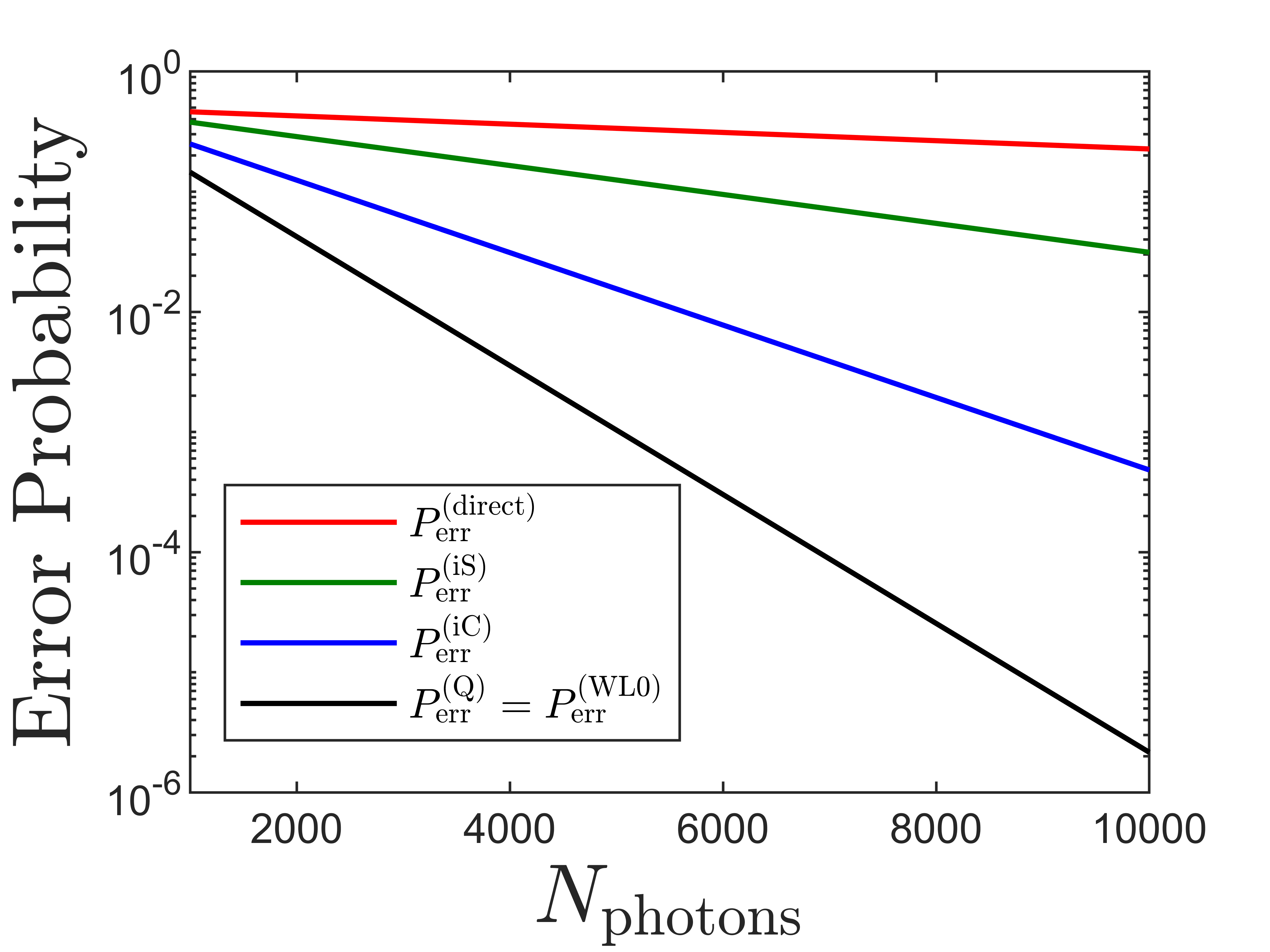}
    \caption{Chernoff scaling of error probabilities in discriminating States A and B for various measurement schemes. $P_\text{err}^\text{(Q)}$ indicates the error probability scaling dictated by the quantum Chernoff bound.}
    \label{fig:weak_decay_path_detection_method_comp}
\end{figure}

The discussion to this point has focused on the specific task of estimating the square-root ratio of time constants given an incoherent sum of two exponential decays of equal likelihood. This exact scenario may be seldom if ever encountered in any real experiment. It serves, however, as a basic example demonstrating the potential power in measurement and inference schemes that more fully exploit the finite temporal coherence available when quantum sources spontaneously emit. Here we present an additional example of this fact by considering a binary hypothesis test in which one aims to discriminate between the following two states: A) a single exponential decay and B) that same single exponential decay incoherently mixed with a second weaker decay rate. To facilitate some quantification, we set the main decay channel's time constant to $\tau_A = 1$ ns and the second channel's constant to $\tau_B = 1.25$ ns. State A is defined by a unit probability of decay via the main channel; State B is defined by a $p_A = 0.9$ probability of decay via the main channel and a $p_B = 0.1$ probability of decay via the second channel. The one-photon density operators representing these two states are:
\begin{subequations}
\begin{equation}
    \rho_A = \ket{\psi_A}\bra{\psi_A}
\end{equation}
\begin{equation}
    \rho_B = p_A\ket{\psi_A}\bra{\psi_A} + p_B\ket{\psi_B}\bra{\psi_B}
\end{equation}
\end{subequations}
with
\begin{subequations}
\begin{equation}
    \ket{\psi_A} = \int_{-\infty}^\infty \mathrm{d}t \, \psi(t;\tau_A) a^\dagger(t) \ket{0}.
\end{equation}
\begin{equation}
    \ket{\psi_B} = \int_{-\infty}^\infty \mathrm{d}t \, \psi(t;\tau_B) a^\dagger(t) \ket{0}.
\end{equation}
\end{subequations}
and $\psi(t;\tau_A)$ and $\psi(t;\tau_B)$ defined as in Eq. (\ref{eq_psi_defn}). If a total of $N_\text{photons}$ identically prepared photons are available for detection before a decision is made, the probability of error in the discrimination task improves exponentially with asymptotic scaling:
\begin{equation}
    P_\text{err} \sim \frac{1}{2} e^{-\xi N_\text{photons}},
\end{equation}
where the factor $\xi$ is given by the Chernoff bound, for which there exist both classical \cite{chernoff1952measure} and quantum \cite{PhysRevLett.98.160501} definitions. The classical Chernoff bound is appropriate when discriminating between two classical probability distributions. For the specific case of discriminating States A and B by implementation of a direct lifetime decay measurement, the task is equivalent to distinguishing between the two classical probability density functions (Fig. \ref{fig:weakdecay_scheme}):
\begin{subequations} \label{eq_weak_cl_dists}
\begin{equation}
    f_A(t) = |\psi(t;\tau_A)|^2
\end{equation}
\begin{equation}
    f_B(t) = p_A|\psi(t;\tau_A)|^2 + p_B|\psi(t;\tau_B)|^2. 
\end{equation}
\end{subequations}
Assuming either state is equally likely, the classical Chernoff bound associated with direct measurement can then be computed according to:
\begin{equation}
    \xi^\text{(direct)} = - \log \min_{0\leq s \leq1} \Big[ \int_0^\infty \mathrm{d}t \, f_A^s(t) f_B^{1-s}(t)\Big].
\end{equation}
Direct measurement is one of infinitely many possible measurement schemes that can distinguish States A and B with some finite probability. A general measurement's ability to discriminate $\rho_A$ and $\rho_B$ (again assuming equal \textit{a priori} likelihood) is governed by the quantum Chernoff bound \cite{PhysRevLett.98.160501}:
\begin{equation}
    \xi^\text{(Q)} = - \log \min_{0\leq s \leq1} \text{Tr}\Big[\rho_A^s \rho_B^{1-s}\Big].
\end{equation}
Since $\rho_A$ is pure in the case under consideration the above expression simplifies to:
\begin{equation}
    \xi^\text{(Q)} = - \log \braket{\psi_A|\rho_B|\psi_A}.
\end{equation}
In such a case it's been established that a projection measurement $\{\rho_A, \mathbb{I}-\rho_A\}$ would saturate the quantum limit to hypothesis discrimination \cite{kargin2005chernoff}. In Fig. \ref{fig:weak_decay_path_detection_method_comp} we plot the approximate probability of error associated with the direct measurement classical Chernoff bound and the quantum Chernoff bound as a function of number of detected photons. Evidently the error rate can be improved by orders of magnitude for a number of photons feasibly detected from a single molecule.
For the same task we also analyzed the performance of measurements based on interferometers iC and iS as described in Section \ref{section_interfs}, but with the substitution $\bar{\tau} \mapsto \tau_A$. We computed the classical Chernoff bounds associated with these two schemes, and plot the associated probabilities of error in Fig. \ref{fig:weak_decay_path_detection_method_comp}. These schemes do not saturate the quantum bound, but do offer significant improvement over direct measurement for sufficiently many photons.

Detection of a weak decay path as described in this section is just one example of a symmetric binary hypothesis test related to spontaneous emission lifetime that one might conceivably encounter. In this case the optimal measurement is straightforward to derive (at least on paper). For more general symmetric binary hypothesis tests in which both possible states are mixed, the optimal measurement scheme is not necessarily clear. Our preliminary calculations indicate that projection measurements onto WL modes with appropriately chosen reference lifetime at least offer significant improvement over direct measurement. This would be in line with the previously reported observation that Hermite-Gaussian projection saturates the quantum Chernoff bound for certain binary discrimination tasks related to sub-diffraction imaging \cite{lu2018quantum,grace2021quantum}. A more thorough investigation of the performance of WL projection measurements for generalized binary discrimination tasks will be the subject of future work.
\section{\label{sec:level1}Conclusion\protect}
In conclusion, we have surveyed several scenarios relevant to estimation, resolution, and discrimination of the lifetime of a quantum optical emitter from the perspective of quantum information theory. While the Fisher information associated with estimation of a single exponential lifetime is saturated by a direct TCSPC measurement, the conventional measurement scheme is fundamentally challenged when presented with statistical mixtures of exponential decays. Namely, we find that the direct-measurement CFI for estimating the separation between two mutually incoherent, highly overlapping decays vanishes as the two decay constants approach one another. In the same limit, however, the QFI remains finite, indicating that a carefully redesigned measurement can potentially uncover a wealth of hidden information. We prove that a projective measurement onto exponentially-weighted Laguerre modes would saturate the QFI, and we speculate on possible routes to experimental realization. Beyond the specific example of bi-exponential resolution, our results suggest that a variety of tasks related to lifetime classification stand to benefit from a more thorough exploitation of the emitted photon's coherence. We analyzed the discrimination of A) a single-exponential decay from B) the same decay mixed with a weak second decay channel by enumerating the quantum and classical Chernoff bounds. 

There remains a significant amount of research to be done in order to assess the experimental relevance of the findings presented here. As a starting point we have invoked a simplistic model for photon emission by a two-level quantum emitter with fixed energy splitting and a variable lifetime. The situation encountered under practical experimental conditions can be much more complicated, especially for single-molecule fluorophores embedded in condensed environments under ambient conditions. The optical spectra of molecular emitters are much richer (even in vacuum) due to coupling to vibrational modes. In reality a change in the environment of an emitter can also cause shifts in energy levels, resulting in a more convoluted spectro-temporal response. In our study we have assumed a fixed and known $\omega_0$. While the metrics we derive do not depend explicitly on $\omega_0$, and while they do not require any multiphoton interference, a degree of spectral homogeneity from photon to photon might nonetheless be important.

Limitations of the physical model aside, our results tie in nicely with recent and ongoing work that demonstrates novel quantum-information-inspired approaches to super-resolving the positions of mutually incoherent emitters \cite{doi:10.1080/00107514.2020.1736375,tsang2016quantum,PhysRevLett.117.190801,Nair:16,Tsang_2017,PhysRevA.95.063847,PhysRevA.97.023830,Rehacek:17,Tang:16,Larson:19,Hassett:18,Zhou:19,PhysRevLett.118.070801,Paur:16,Paur:18,Paur:19} and the temporal offsets of ultrafast pulses \cite{PhysRevLett.121.090501,PhysRevResearch.3.033082,PRXQuantum.2.010301}. That the lifetime equivalent of ``Rayleigh's Curse'' can be overcome by projection onto a particular set of weighted orthogonal polynomials is an intriguing generalization of the previous observation that projection onto a different set of weighted orthogonal polynomials does the same for the spatial ``Curse'' \cite{tsang2016quantum,Rehacek:17}. Just as preliminary work describing the resolution of exactly two point emitters has led to recent consideration of more general imaging scenarios \cite{PhysRevA.99.012305,PhysRevA.99.013808,Bisketzi_2019,PhysRevLett.124.080503}, resolution and discrimination of more complex lifetime mixtures will be the subject of future investigation. Finally, the work presented here is one of a few recent studies that highlights the utility in generalizing the now-familiar information-theoretic tasks of single-molecule microscopy to the realm of quantum information theory \cite{PhysRevLett.121.023904,PhysRevResearch.2.033114,zhang2021singleI,zhang2021singleII}. Given the preciousness of photons and the zoo of parameters commonly estimated and classified in single-molecule microscopy, we anticipate this will continue to be a fruitful line of inquiry. Further research in this vein is underway.
\begin{acknowledgments}
We thank Yunkai Wang for helpful comments on the manuscript.
\end{acknowledgments}
\appendix
\section{Derivation of Eq. \ref{eq_QFI_resolution}} \label{appendix_A}
Equation (\ref{eq_QFI_defn}) defines the elements of the QFI matrix in terms of the SLD, while Eq. (\ref{eq_SLD_implicit}) gives an implicit relation between the SLD and the density operator $\rho$. To compute the SLDs $\mathcal{L}_{\bar{\tau}}$ and $\mathcal{L}_{\varepsilon}$ given the state defined in Eq. (\ref{eq_rho_resolution}) we make use of the following explicit relation:
\begin{equation} \label{eq_SLD_explcit}
    \mathcal{L}_i = \sum_{k,l; D_k+D_l \neq 0} \frac{2}{D_k + D_l} \braket{e_k|\partial_{\theta_i}\rho|e_l} \ket{e_l}\bra{e_k}
\end{equation}
where $\{\ket{e_k}\}$ are orthonormal eigenvectors of the operator $\rho$ with associated eigenvalues $\{D_k\}$ such that:
\begin{equation}
    \rho = \sum_k D_k \ket{e_k}\bra{e_k}.
\end{equation}
While the relevant Hilbert space is infinite, we need only consider as many $\ket{e_k}$ as it takes to provide a basis for $\text{span} \{ \ket{\psi_0},\ket{\psi_1},\ket{\partial_{\bar{\tau}}\psi_0},\ket{\partial_{\bar{\tau}}\psi_1},\ket{\partial_\varepsilon\psi_0},\ket{\partial_\varepsilon\psi_1} \}$. In this case four such vectors constitute such a basis. The states $\ket{e_1}$ and $\ket{e_2}$ can be arrived at by inspection:
\begin{subequations}
\begin{equation}
    \ket{e_1} = \frac{1}{\sqrt{2(1+\chi)}}\left( \ket{\psi_0} + \ket{\psi_1} \right)
\end{equation}
\begin{equation}
    \ket{e_2} = \frac{1}{\sqrt{2(1-\chi)}}\left( \ket{\psi_0} - \ket{\psi_1} \right),
\end{equation}
\end{subequations}
for which
\begin{subequations}
\begin{equation}
    D_1 = \frac{1+\chi}{2}
\end{equation}
\begin{equation}
    D_2 = \frac{1-\chi}{2}
\end{equation}
\end{subequations}
and
\begin{equation}
    \chi = \braket{\psi_0|\psi_1}.
\end{equation}
Next $\ket{e_3}$ and $\ket{e_4}$ can be found via a modified Gram-Schmidt process. First we define:
\begin{subequations}
\begin{equation}
    \ket{v_3} = \ket{\partial_{\tau_0}\psi_0} - \braket{e_1|\partial_{\tau_0}\psi_0}\ket{e_1} - \braket{e_2|\partial_{\tau_0}\psi_0}\ket{e_2}
\end{equation}
\begin{equation}
    \ket{v_4} = \ket{\partial_{\tau_1}\psi_1} - \braket{e_1|\partial_{\tau_1}\psi_1}\ket{e_1} - \braket{e_2|\partial_{\tau_1}\psi_1}\ket{e_2}
\end{equation}
\end{subequations}
and normalize to give:
\begin{subequations}
\begin{equation}
    \ket{f_3} = \frac{\ket{v_3}}{\|v_3\|}
\end{equation}
\begin{equation}
    \ket{f_4} = \frac{\ket{v_4}}{\|v_4\|}.
\end{equation}
\end{subequations}
Letting $\Upsilon$ be defined by
\begin{equation}
    \Upsilon = \braket{f_3|f_4},
\end{equation}
we then arrive at 
\begin{subequations}
\begin{equation}
    \ket{e_3} = \frac{1}{\sqrt{2(1+\Upsilon)}}\left( \ket{f_3} + \ket{f_4} \right)
\end{equation}
\begin{equation}
    \ket{e_4} = \frac{1}{\sqrt{2(1-\Upsilon)}}\left( \ket{f_3} - \ket{f_4} \right),
\end{equation}
\end{subequations}
for which
\begin{equation}
    D_3 = D_4 = 0.
\end{equation}
With $\{\ket{e_1},\ket{e_2},\ket{e_3},\ket{e_4}\}$ and $\{D_1,D_2,D_3,D_4\}$ now in hand, $\mathcal{L}_{\bar{\tau}}$ and $\mathcal{L}_\varepsilon$ can be computed directly via Eq. (\ref{eq_SLD_explcit}). Finally, $\mathcal{K}$ can be computed via Eq. (\ref{eq_QFI_defn}), yielding the result of Eq. (\ref{eq_QFI_resolution}). 

\section{Proof of Eq. (\ref{eq_iFT_WL_relation})} \label{appendix_B}
\begin{figure}
    \centering
    \includegraphics{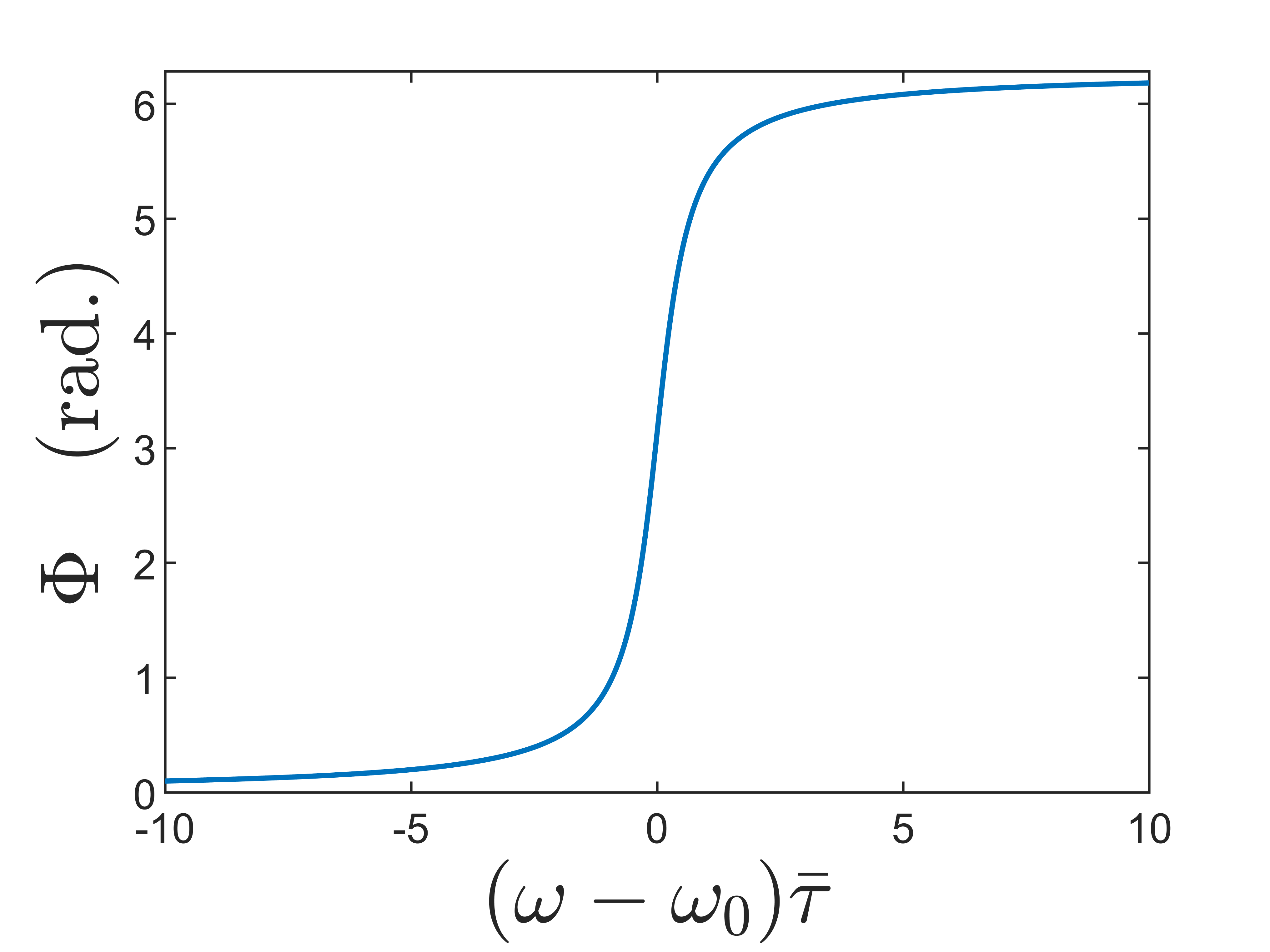}
    \caption{Phase angle $\Phi$ as defined in Eq. (\ref{eq_Phi}). We apply $2 \pi$ phase-wrapping to enforce continuity.}
    \label{fig:Phi_plot}
\end{figure}
Equations (\ref{eq_phi_defn}) and (\ref{eq_define_phi_tilde}) can be combined to give:
\begin{equation}
    \tilde{\phi}_n(\omega) = \frac{1}{\sqrt{2\pi \bar{\tau}}} \int_0^\infty \mathrm{d}t \, e^{i[(\omega-\omega_0) + i/(2\bar{\tau})]t} L_n\left(\frac{t}{\bar{\tau}}\right).
\end{equation}
Change of variables to the unitless real number
\begin{equation}
    T = t/\bar{\tau}
\end{equation}
and unitless complex number
\begin{equation} \label{eq_xi_defn}
    \zeta = (\omega-\omega_0)\bar{\tau} + \frac{i}{2}
\end{equation}
leads to the more compact expression:
\begin{equation}
        \tilde{\phi}_n(\omega) = \sqrt{\frac{\bar{\tau}}{2\pi}} \int_0^\infty \mathrm{d}T \, e^{i\zeta T} L_n(T).
\end{equation}
We will prove by induction that the relation
\begin{equation} \label{eq_recursion_xi}
    \tilde{\phi}_n(\omega) = \frac{i}{\zeta}\sqrt{\frac{\bar{\tau}}{2\pi}}\left(\frac{1+i\zeta}{i\zeta} \right)^n
\end{equation}
holds for all $n\geq0$. We begin with the base case of $n=0$. The integral
\begin{equation}
    \tilde{\phi}_0(\omega) = \sqrt{\frac{\bar{\tau}}{2\pi}} \int_0^\infty \mathrm{d}T \, e^{i\zeta T}
\end{equation}
can be computed directly to give:
\begin{equation}
    \tilde{\phi}_0(\omega) = \frac{i}{\zeta}\sqrt{\frac{\bar{\tau}}{2\pi}}.
\end{equation}
Thus Eq. (\ref{eq_recursion_xi}) indeed holds in the base case of $n=0$.

Next we assume that
\begin{equation} \label{eq_assumption}
    \tilde{\phi}_k(\omega) = \frac{i}{\zeta}\sqrt{\frac{\bar{\tau}}{2\pi}}\left(\frac{1+i\zeta}{i\zeta} \right)^k
\end{equation}
holds $\forall k \in \{0,...,n-1\}$ and consider $\tilde{\phi}_n(\omega)$:
\begin{equation}
        \tilde{\phi}_n(\omega) = \sqrt{\frac{\bar{\tau}}{2\pi}} \int_0^\infty \mathrm{d}T \, e^{i\zeta T} L_n(T).
\end{equation} 
We integrate by parts to give:
\begin{eqnarray} \label{eq_intbyparts1}
    \tilde{\phi}_n(\omega) = \sqrt{\frac{\bar{\tau}}{2\pi}}\Bigg\{&& \frac{1}{i\zeta}\left[L_n(T)e^{i\zeta T}\right]\Big|_0^\infty \nonumber \\ &&-\frac{1}{i\zeta} \int_0^\infty \mathrm{d}T \, L^\prime_n(T) e^{i\zeta T} \Bigg\}.
\end{eqnarray}
Noting that $L_n(0)=1 \, \forall n$ and that
\begin{equation}
    \lim_{T \to \infty}\left[ L_n(T)e^{i\zeta T} \right]=0
\end{equation}
due to the decaying exponential portion simplifies Eq. (\ref{eq_intbyparts1}) to:
\begin{equation} \label{eq_intbyparts_simp}
    \tilde{\phi}_n(\omega) = \frac{i}{\zeta}\sqrt{\frac{\bar{\tau}}{2\pi}}\Bigg\{1+\int_0^\infty \mathrm{d}T \, L^\prime_n(T) e^{i\zeta T} \Bigg\}.
\end{equation}
From the well-known Rodrigues equation for Laguerre polynomials \cite{abramowitz1964handbook} one can deduce the recursion relation:
\begin{equation}
    \frac{d}{dT} L_n(T) = \left( \frac{d}{dT} - 1\right) L_{n-1}(T),
\end{equation}
for $n>0$, which can be applied repeatedly to rewrite Eq. (\ref{eq_intbyparts_simp}):
\begin{eqnarray}
    \tilde{\phi}_n(\omega) &&= \frac{i}{\zeta}\sqrt{\frac{\bar{\tau}}{2\pi}}\Bigg\{1-\sum_{k=0}^{n-1}\int_0^\infty \mathrm{d}T \, L_k(T) e^{i\zeta T} \Bigg\} \nonumber \\ &&= \frac{i}{\zeta}\sqrt{\frac{\bar{\tau}}{2\pi}}\Bigg\{1-\sqrt{\frac{2\pi}{\bar{\tau}}}\sum_{k=0}^{n-1} \tilde{\phi}_k(\omega) \Bigg\}.
\end{eqnarray}
Invoking our assumption in Eq. (\ref{eq_assumption}):
\begin{equation}
    \tilde{\phi}_n(\omega) = \frac{i}{\zeta}\sqrt{\frac{\bar{\tau}}{2\pi}}\Bigg\{1-\frac{i}{\zeta}\sum_{k=0}^{n-1} \left[\frac{1+i\zeta}{i\zeta}\right]^k \Bigg\}.
\end{equation}
The geometric series in the above equation can be substituted by its closed-form solution:
\begin{equation}
    \tilde{\phi}_n(\omega) = \frac{i}{\zeta}\sqrt{\frac{\bar{\tau}}{2\pi}}\left\{1-\frac{i}{\zeta} \left[\frac{1-\left(\frac{1+i\zeta}{i\zeta}\right)^n}{1-\left(\frac{1+i\zeta}{i\zeta}\right)}\right] \right\},
\end{equation}
which can be simplified to yield Eq. (\ref{eq_recursion_xi}), thus concluding this portion of the proof.

We end this section by establishing the relationship between Eq. (\ref{eq_recursion_xi}) and Eq. (\ref{eq_iFT_WL_relation}). From Eq. (\ref{eq_recursion_xi}) and the definition in Eq. (\ref{eq_xi_defn}) follows:
\begin{equation} \label{eq_almostthere}
    \tilde{\phi}_n(\omega) = \frac{i}{(\omega-\omega_0)\bar{\tau} + i/2}\sqrt{\frac{\bar{\tau}}{2\pi}} \left\{\frac{(\omega-\omega_0)\bar{\tau}-i/2}{(\omega-\omega_0)\bar{\tau}+i/2}\right\}^n.
\end{equation}
The complex number enclosed in curly brackets in Eq. (\ref{eq_almostthere}) clearly has unit magnitude and so can be written:
\begin{equation}
    \frac{(\omega-\omega_0)\bar{\tau}-i/2}{(\omega-\omega_0)\bar{\tau}+i/2} = e^{i\Phi}
\end{equation}
for some real number $\Phi$. The phase function is given explicitly by:
\begin{equation} \label{eq_Phi}
    \Phi(\omega-\omega_0;\bar{\tau}) = \tan^{-1}{\left[ \frac{-(\omega-\omega_0)\bar{\tau}}{(\omega-\omega_0)^2\bar{\tau}^2 - 1/4} \right]}.
\end{equation}
Equation (\ref{eq_iFT_WL_relation}) follows directly. The function $\Phi(\omega-\omega_0;\bar{\tau})$ is plotted in Fig. \ref{fig:Phi_plot}.
\section{Additional supporting figures} \label{appendix_C}
\begin{figure}
    \centering
    \includegraphics{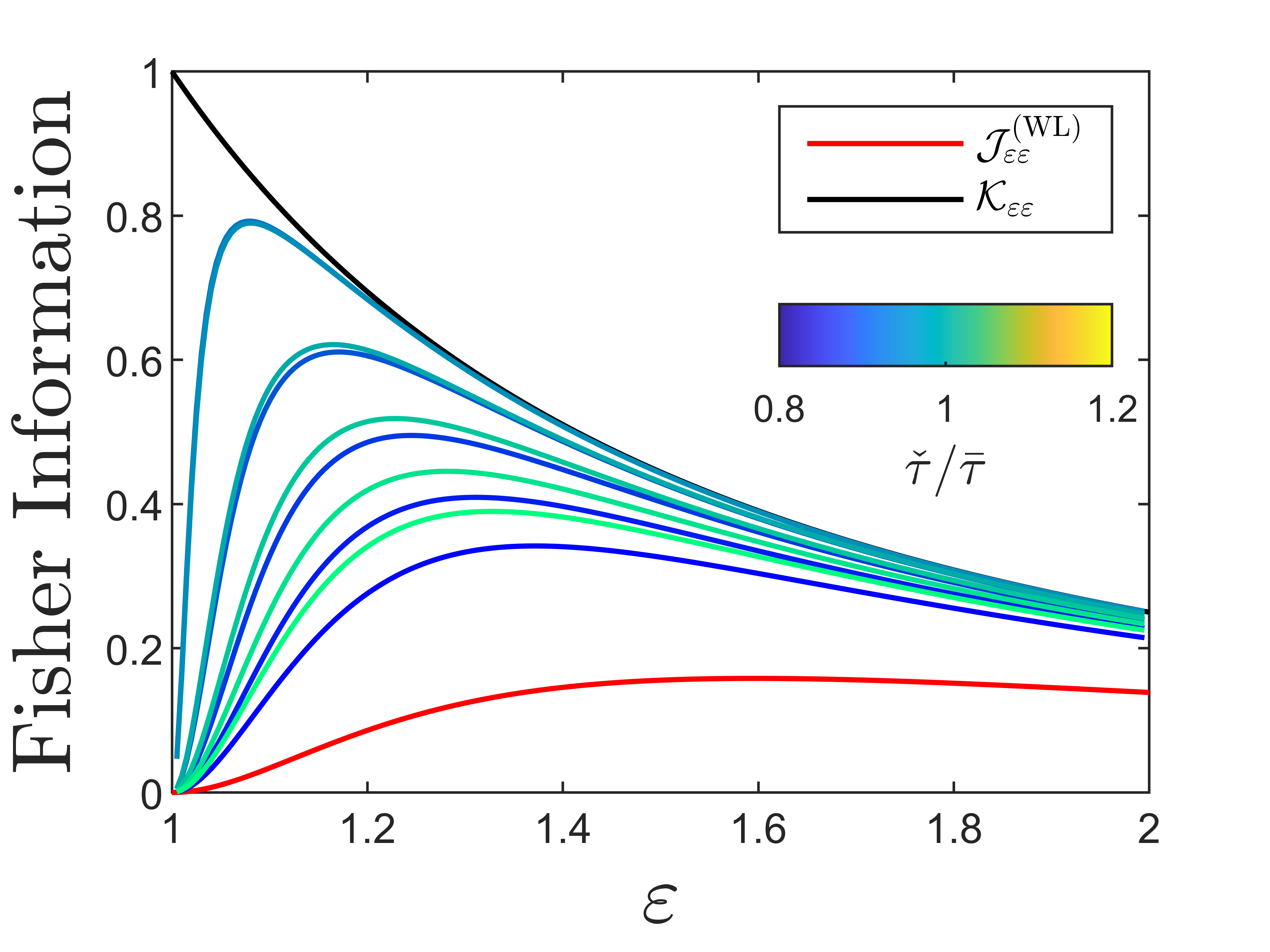}
    \caption{Fisher information for $\varepsilon$ estimation by projection onto shifted WL modes $\{\phi_n(t;\check{\tau})\}_n$ with various mismatched mean lifetimes $\check{\tau} \neq \bar{\tau}$. Similar to the performance of centroid-mismatched SPADE vis-\`a-vis spatial resolution \cite{tsang2016quantum}, the effect of mismatch is a diminished (and eventually vanishing) CFI at sufficiently small $\varepsilon$. Importantly, bounding the mismatch below some small but finite threshold leaves a considerable range for which $\mathcal{J}^\text{(WL)}_{\varepsilon\varepsilon} \gg \mathcal{J}^\text{(direct)}_{\varepsilon\varepsilon}$.}
    \label{fig:taubar_mismatch}
\end{figure}
\begin{figure}
    \centering
    \includegraphics{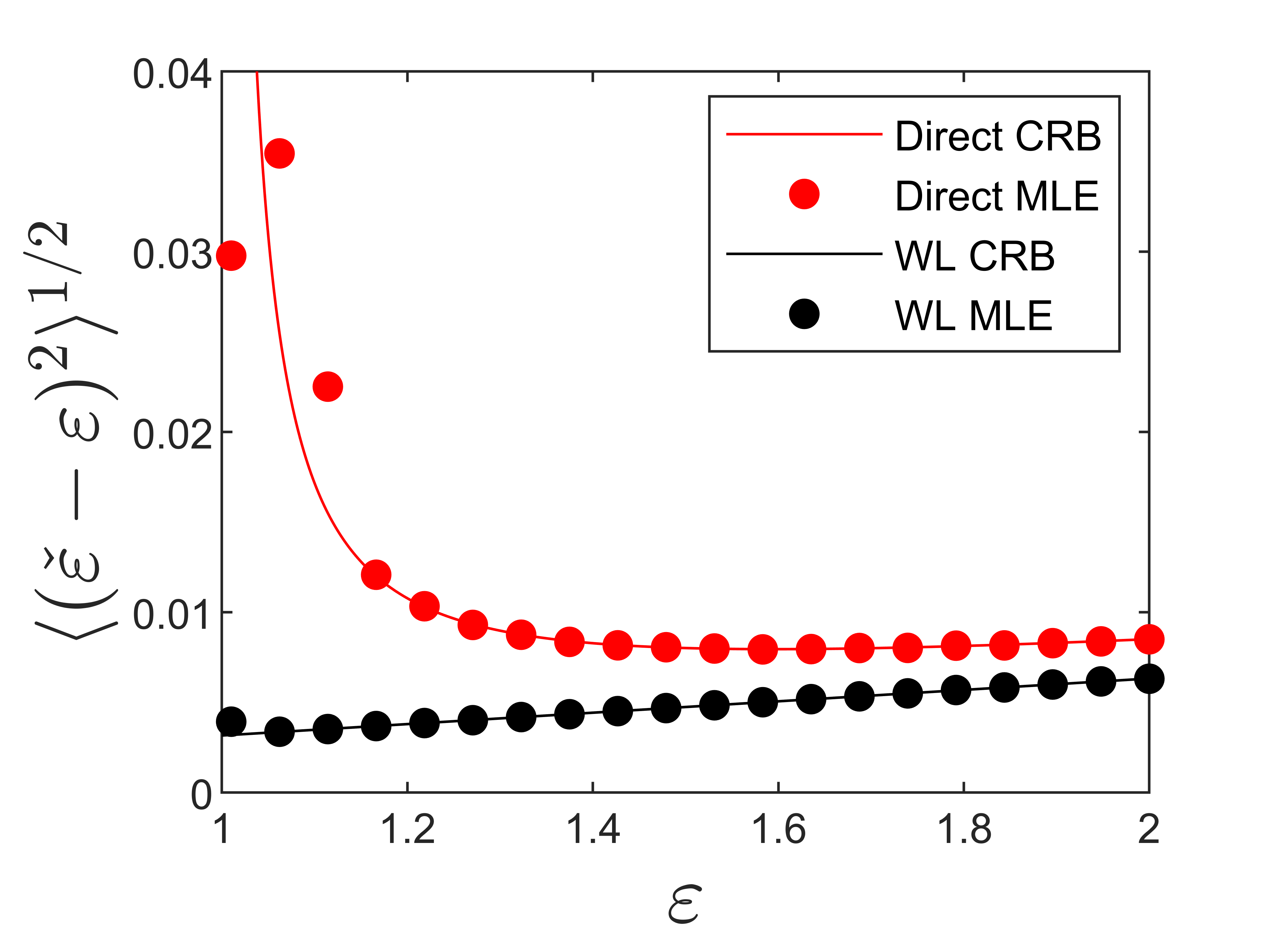}
    \caption{Monte Carlo simulation of bi-exponential resolution task, with $N_\text{photons} = 10^5$ and results averaged over $10^5$ trials. Here we fixed $\bar{\tau} = 1$ ns and varied $\varepsilon$. At each $\varepsilon$ we generated random data corresponding to a direct measurement and a WL projection measurement, then used the maximum likelihood estimator (MLE) appropriate for each modality to produce estimates $\check{\varepsilon}$. The root-mean-square-errors (RMSE) of both MLEs are plotted. We also plot the CRBs associated with direct measurement and WL projection for comparison. For sufficiently large $\varepsilon$ the RMSEs are bounded by their respective CRBs, as must be the case for any unbiased estimator. At very small $\varepsilon$ the RMSE of the MLE deviates from the CRB due to the onset of bias in the estimators. Nonetheless, the WL projection significantly outperforms direct measurement at small $\varepsilon$.}
    \label{fig:monte_carlo}
\end{figure}
Figure \ref{fig:taubar_mismatch} shows the effect of mean-lifetime mismatch on $\mathcal{J}^\text{(WL)}_{\varepsilon\varepsilon}$. Results of a Monte Carlo simulation of the bi-exponential resolution task are given in Fig. \ref{fig:monte_carlo}.
\bibliography{mybib.bib}
\end{document}